\documentclass[aps,pra,twocolumn,groupedaddress,showpacs]{revtex4}
\usepackage{dcolumn}
\usepackage{bm}
\usepackage{graphicx}
\usepackage{amsmath}
\usepackage{amssymb}
\usepackage{xcolor}
\newcommand*\bra[1]{\langle#1\vert}
\newcommand*\ket[1]{\vert#1\rangle}
\newcommand*\braket[2]{\langle#1\vert#2\rangle}
\newcommand*\braopket[3]{\langle#1\vert#2\vert#3\rangle}
\newcommand*\ip{I_{\rm p}}
\newcommand*\up{U_{\rm p}}
\newcommand*\Ai{{\mathop{\mathrm{Ai}}}}
\newcommand*\bfa{\mathbf{a}}
\newcommand*\bfb{\mathbf{b}}
\newcommand*\bfm{\mathbf{m}}
\newcommand*\bfn{\mathbf{n}}

\newcommand*\bfA{\mathbf{A}}
\newcommand*\bfB{\mathbf{B}}

\newcommand*\bfP{\mathbf{P}}

\newcommand*\bfS{\mathbf{S}}

\newcommand*\bfU{\mathbf{U}}

\newcommand*\bfzero{\mathbf{0}}
\newcommand*\bfone{\mathbf{1}}
\newcommand*\I{\mathrm{i}}
\newcommand*\E{\mathrm{e}}
\newcommand*\rms{\mathrm{s}}
\newcommand*\rmC{\mathrm{C}}
\newcommand*\rmD{\mathrm{D}}
\newcommand*\rmI{\mathrm{I}}
\newcommand*\rmL{\mathrm{L}}
\newcommand*\rmN{\mathrm{N}}
\newcommand*\rmT{\mathrm{T}}
\newcommand*\rmV{\mathrm{V}}
\newcommand*\rmX{\mathrm{X}}
\newcommand*\vecd{\vec{d}}
\newcommand*\veck{\vec{k}}
\newcommand*\vecr{\vec{r}}
\newcommand*\vecq{\vec{q}}
\newcommand*\vecA{\vec{A}}

\newcommand*\vecF{\vec{F}}

\newcommand*\vecR{\vec{R}}
\renewcommand{\arraystretch}{1.3}

\begin{document}

\title{Full-dimensional treatment of short-time vibronic dynamics in molecular high-harmonics generation process in methane}

\author{Serguei Patchkovskii}
\email[]{Serguei.Patchkovskii@mbi-berlin.de}
\affiliation{Max-Born Institute, Max-Born-Stra{\ss}e 2A, 12489 Berlin, Germany}

\author{Michael S. Schuurman}
\affiliation{Department of Chemistry and Biomolecular Sciences, University of Ottawa, D'Iorio Hall, 10 Marie Curie, Ottawa ON Canada,K1N 6N5}
\affiliation{ational Research Council of Canada, 100 Sussex Dr., Ottawa, Ontario, Canada, K1A 0A6}

\date{\today}

\begin{abstract}
We present derivation and implementation of the Multi-Configurational Strong-Field
Approximation with Gaussian nuclear Wave Packets (MC-SFA-GWP) -- a version of
the molecular strong-field approximation which treats all electronic and
nuclear degrees of freedom, including their correlations, quantum-mechanically.
The technique allows, for the first time, realistic simulation of high-harmonic
emission in polyatomic molecules without invoking reduced-dimensionality models
for the nuclear motion or the electronic structure. We use MC-SFA-GWP to model
isotope effects in high-harmonics generation (HHG) spectroscopy of methane.
The HHG emission in this molecule transiently involves strongly
vibronically-coupled $^2F_2$ electronic state of the $\rm CH_4^+$ cation.  We
show that the isotopic HHG ratio in methane contains signatures of: a)
field-free vibronic dynamics at the conical intersection (CI); b) resonant
features in the recombination cross-sections; c) laser-driven bound-state
dynamics; as well as d) the well-known short-time Gaussian decay of the
emission. We assign the intrinsic vibronic feature (a) to a relatively
long-lived ($\ge4$~fs) vibronic wave packet of the singly-excited $\nu_4$
($t_2$) and $\nu_2$ ($e$) vibrational modes, strongly coupled to the components
of the $^2F_2$ electronic state.  We demonstrate that these physical effects
differ in their dependence on the wavelength, intensity, and duration of the
driving pulse, allowing them to be disentangled.  We thus show that HHG
spectroscopy provides a versatile tool for exploring both conical intersections
and resonant features in photorecombination matrix elements in the regime not
easily accessible with other techniques. 
\end{abstract}


\maketitle


\section{Introduction}

One of the elusive targets being pursued by the rapidly-developing research
area of strong-field and attosecond science is following the electronic and
nuclear motion in atoms and molecules on their natural, atto- and femto-second
time
scales\cite{Gouliemakis07a,Krausz09a,Salieres12a,Lepine14a,Calegari16a,Ramasesha16a}.
High-harmonics spectroscopy (HHS)\cite{Haessler11a}, based on the celebrated
high-harmonics generation (HHG) process\cite{Ferray88a,Corkum93a}, is one of
the most powerful and versatile tools in the arsenal of the attosecond science.
Among other examples\cite{Kanai08b,Fleischer14a}, it has been successfully
employed to image the ``molecular
orbitals''\cite{Itatani04a,Haessler10a,Salieres12a,Schwarz06a} and resolve
multiple final states in strong-field ionization\cite{Farrell11a,Ferre15a};
follow evolution of a bound
electronic\cite{Smirnova09a,Smirnova09d,Mairesse10a,Torres10a,Diveki12a,
Leeuwenburgh13a,Kim13a,Kraus15a,Bruner16a},
rotational\cite{Wagner06a,Mairesse08a,Spector14a} and
nuclear\cite{Li08a,Ferre15a} wave packets; monitor electron correlation in
atoms\cite{Shiner11a}; measure molecular
chirality\cite{Cireasa15a,Smirnova15a}; resolve the time tunneling electron
emerges from underneath the barrier\cite{Shafir12a}; probe strongly-driven
electrons in the continuum\cite{Zhou08a,Smirnova09d,Worner09a,Orenstein17a};
and control attosecond emission from molecules\cite{Boutu08a,Zhou09a}.

HHG is a highly non-linear process\cite{Corkum93a,Lewenstein94a}, which often
makes the interpretation of experimental results complex and controversial.
Similar to conventional spectroscopies, measurement of the isotope effects and
quantum-path interferences in HHG has emerged as one of the most helpful
techniques in the detailed analysis of high-harmonics
spectra\cite{Lein05a,Baker06a,Marangos08a,Baker08a,Kanai08a,Haessler09a,
Mizutani11a,Farrell11a,Kraus13a,Zair13a,Lan17a}. In the form of the PACER
(Probing Attosecond Dynamics by Chirp-Encoded Recollision\cite{Baker07a})
experiment, isotope-dependent HHS promises direct access to the electronic and
nuclear dynamics on a few- and sub-femtosecond time scale, including
measurement of the phases of the nuclear
wave packets\cite{Kanai08a,Haessler09a}.

For the electronically simple cases, where the Born-Oppenheimer (BO) variable
separation applies, the theory underlying PACER experiments is well
understood\cite{Lein05a,Lein07a}. The relevant theory object, the short-time
nuclear autocorrelation function\cite{Lein05a}, is readily available for most
small molecules\cite{Patchkovskii09a}. At the same time, the BO assumptions
fail already for one of the first molecules examined with PACER spectroscopy.
In methane, $\rm CH_4$, the key dynamics occur in the vicinity of a
symmetry-required triple electronic surface intersection in a
transiently-prepared $\rm CH_4^+$ cation. In larger, more electronically
complex systems, which are beginning to be explored with
HHS\cite{Ganeev09a,Wong10a,Torres10b,Vozzi10a,Wong11a,Wong13a,Hutchison13a,Alharbi15a}
(See Ref.~\cite{Marangos16a} for a review), conical intersections and
non-separable vibronic dynamics are expected to become
ubiquitous\cite{Yarkony04a,Yarkony12a}. 

The effects of the non-Born-Oppenheimer dynamics and isotope effects in HHG
have been studied for more than 20 years\cite{Chelkowski96a}. Nonetheless, most
of the non-BO numerical studies of HHG still concern with the dynamics of
simple, one- and two- electron diatomics ($\rm
H_2$\cite{Chirila08a,Bandrauk09a,Chirila09b,Daniele09a}, $\rm
H_2^+$\cite{Chelkowski96a,Bandrauk08a,Zhao08a,Bandrauk09a,Daniele09a,Castiglia11a,
Bredtmann12a,Zheng12a, Nguyen13a,Ahmadi14a,Bian14a,Morales14a,Ahmadi16a,
Lara-Astiaso16a,Li17a}, $\rm HeH^{2+}$\cite{Miao13a}), usually with the
additional approximations of the reduced
dimensionality\cite{Bandrauk08a,Zhao08a,Bandrauk09a,Chirila09b,Daniele09a,
Castiglia11a,Bredtmann12a, Zheng12a,Miao13a,Nguyen13a,Morales14a,Li17a} or
special field
polarizations\cite{Chelkowski96a,Ahmadi14a,Bian14a,Ahmadi16a,Lara-Astiaso16a}.
Some attempts at extending the theory of nuclear dynamics in HHG to more
general non-BO systems have been reported in the
literature\cite{Walters07a,Chirila08a,Walters09a,Faisal10a,Falge10a,Madsen10a,
Lisinetskaya11a,Le12a,Forster13a,Patchkovskii14a,Mondal14a,Mondal15a,Rao15a,
Rao15b,Rao15c,Vacher15a,Arnold17a}.  Nonetheless, no fully satisfactory
solution to this problem, which requires simultaneous consideration of
non-perturbative continuum electronic dynamics, vibronic dynamics, and their
correlations, has become available so far.

The goal of this work is to extend the strong-field approximation
(SFA)\cite{Lewenstein94a,Ivanov96a,Becker05a}, a non-perturbative theory
underlying much of attosecond science\cite{Popruzhenko14a,Krausz09a}, to
account for the short-time vibronic dynamics. The rest of this work is
structured as follows. Section~\ref{sec:theory} derives the working expressions
for the Multi-Configurational Strong-Field Approximation with Gaussian nuclear
Wave Packets (MC-SFA-GWP). Section~\ref{sec:details} gives technical details for
the MC-SFA-GWP calculations of HHG emission in methane and deuterated methane,
including the details of the electronic structure calculations, the
diabatization procedure, and integration of the MC-SFA-GWP equations.
Section~\ref{sec:results} presents the results of the numerical calculations.
Specifically, section~\ref{sec:results:auto} discusses vibronic nuclear
autocorrelation function in methane and analyses its features in terms of the
wave packet composition and dynamics. Section~\ref{sec:results:hhg} presents
calculated HHG spectra for three wavelengths ($800$, $1600$, and $2400$~nm) and
two intensities ($300$ and $1000$ TW~cm$^{-2}$) of the driving field, and
dissects the physical origin of the observed spectral features.
Section~\ref{sec:results:isotope} analyses calculated PACER signals, compares
them to the available experimental data, and makes predictions for the
previously unexplored range of the experimental parameters.
Section~\ref{sec:conclusions} summarizes the results and offers an outlook for
future developments. Finally, some of the technical aspects of the MC-SFA-GWP
derivation and implementation are relegated to the
appendices~\ref{sec:appendix-A}--\ref{sec:appendix-C}.

\section{Theory\label{sec:theory}}

Our aim is to model high-harmonics emission due to the short-time, coupled
electronic and nuclear dynamics of a molecular system under the influence of an
intense, long-wavelength (near- to mid-IR) laser field.  It is convenient to
write the total Hamiltonian $\hat{H}_\rmT$ as a sum of three terms, all of
which may be time-dependent\cite{Smirnova07b,Smirnova13a,Smirnova13b}:
\begin{align}
  \hat{H}_\rmT\left(t\right) & = \hat{P}_0 \hat{H}_0\left(t\right) \hat{P}_0 
                               + \hat{P}_1 \hat{H}_1\left(t\right) \hat{P}_1 
                               + \hat{P}_1 \hat{V}_\rmL\left(t\right) \hat{P}_0, \label{eqn:hamiltonian}
\end{align}
where $\hat{P}_0$ and $\hat{P}_1$ are respectively projectors onto the neutral
and singly-ionized spaces.  The contribution $\hat{H}_0\left(t\right)$ acts
within the subspace of the neutral states.  The term $\hat{H}_1\left(t\right)$
acts within the subspace of singly-ionized states, while the interaction
Hamiltonian $\hat{V}_\rmL$ describes laser-induced transitions between the two
spaces. In the dipole approximation and length gauge:
\begin{align}
  \hat{V}_\rmL\left(t\right) & = e \vecF\!\left(t\right) \cdot \sum_{i}^{N} \vec{r}_i, \label{eqn:laser}
\end{align}
where $\vec{r}_i$ are coordinates of the $i$-th electron and
$\vecF\!\left(t\right)$ is the laser electric field, which is is assumed to be
slowly-varying. $N$ is the total number of electrons in the system.  In
Eq.~\ref{eqn:hamiltonian}, we have already neglected the possibility of
multiple ionization, as well as the stimulated recombination process.

A formally exact solution of the time-dependent Schr\"{o}dinger equation with
Hamiltonian \ref{eqn:hamiltonian} is given by\cite{Becker05a,Smirnova13a,Smirnova13b}:
\begin{widetext}
\begin{align}
  \Psi\left(\vecr,\vecq,t\right) & = \hat{U}_0\left(t,t_0\right) \Psi\left(\vecr,\vecq,t_0\right)
     - \frac{\I}{\hbar} \int_{t_0}^{t} d t_1 \hat{U}_\rmT\left(t,t_1\right) \hat{V}_\rmL\left(t_1\right) 
       \hat{U}_0\left(t_1,t_0\right) \Psi\left(\vec{r},\vec{q},t_0\right).
     \label{eqn:wavefunction-exact}
\end{align}
\end{widetext}
The propagators $\hat{U}_0$, $\hat{U}_1$, and $\hat{U}_\rmT$ are:
\begin{align}
  \hat{U}_{\rmX}\left(t_j,t_i\right) & = \exp\left(-\frac{\I}{\hbar}\int_{t_i}^{t_j} d t' 
                                              \hat{H}_{\rmX}\left(t'\right) \right), \label{eqn:propagators}
\end{align}
where $\rmX=0,1,\rmT$.
In Eq.~\ref{eqn:wavefunction-exact}, $\vecr$ and $\vecq$ are respectively
electronic and nuclear coordinates.  We assume that the initial wavefunction at
time $t_0$ ($\Psi\left(\vec{r},\vec{q},t_0\right)$) corresponds to the neutral
species.

Classically, the high-harmonics intensity is determined by the power spectrum
of the time-dependent dipole $\vecd\left(t\right)$. Neglecting contributions
arising entirely within the neutral and cation manifolds, the radiating dipole
is given by:
\begin{widetext}
\begin{align}
 \vecd\left(t\right) & = -\frac{\I}{\hbar} \int d\vecr \int d\vecq \int_{t_0}^{t} d t_1
    \left[ \hat{U}_0\left(t,t_0\right) \Psi\left(\vecr,\vecq,t_0\right) \right]^* 
    \hat{V}_\rmD 
    \hat{U}_\rmT\left(t,t_1\right) \hat{V}_\rmL\left(t_1\right) 
    \hat{U}_0\left(t_1,t_0\right) \Psi\left(\vec{r},\vec{q},t_0\right)
 + \hbox{\rm c.c.}, \label{eqn:dipole}
\end{align}
\end{widetext}
where the recombination dipole operator is:
\begin{align}
  \hat{V}_\rmD = e \sum_i^N \vec{r}_i,
\end{align}
with sum running over coordinates of all electrons.

Equation \ref{eqn:dipole} can be brought into a computationally tractable form
by introducing an identity operator $\hat{\bfone}$, given by a sum of identity
operators within the neutral, singly-ionized, etc. spaces:
\begin{align}
  \hat{\bfone}   & = \hat{\bfone}_0 + \hat{\bfone}_1 + \cdots, \label{eqn:identity} \\
  \hat{\bfone}_0 & = \sum_{a\bfn} \ket{\Phi_a} \bra{\Phi_a} \ket{\bfn}\bra{\bfn}, \label{eqn:identity0} \\
  \hat{\bfone}_1 & = \sum_{b\bfm} \ket{\bfm}\bra{\bfm} 
                     \int d \veck \left[ \hat{A}_1 \ket{X_b} \ket{\veck_b} \right] 
                            \left[ \hat{A}_1 \bra{X_b} \bra{\veck_b} \right]. \label{eqn:identity1}
\end{align}

The identity operator within the neutral space ($\hat{\bfone}_0$) is defined in
terms of the neutral electronic states $\ket{\Phi_a}$ (which depend on $\vecr$
and parametrically on $\vecq$) and harmonic nuclear vibrational states
$\ket{\bfn}$ (which depend only on $\vecq$). Quantity $\bfn$ is a vector of
non-negative integers, with each element $\bfn_k$ defining the excitation level
of the $k$-th normal mode of a reference potential $v_{\rm
ref}\left(\vecq\right)$, which is chosen for computational
convenience\cite{Patchkovskii09a,Patchkovskii14a}. 

The singly-ionized space identity operator ($\hat{\bfone}_1$) contains bound
states of the $(N-1)$-electron ion $\ket{X_b}$ and the corresponding
one-electron scattering states $\ket{\veck_b}$ with the asymptotic momentum
$\veck_b$. Functions $\ket{\veck_b}$ depend on the coordinate of the $N$-th
electron $\vecr_N$. Formally, sum over $b$ in Eq.~\ref{eqn:identity1} does
not include autoionizing states $\ket{X_b}$, which belong to the doubly-ionized
continuum. Such states are unlikely to be populated by strong-field ionization,
but may become important due to electron-correlation effects in the
recombination step\cite{Amusia75a,Shiner11a,Patchkovskii12a}. If the lifetime
of such autoionizing state is long compared to the laser cycle, its effects
can still be accounted for, by including the state in Eq.~\ref{eqn:identity1}.

Both $\ket{X_b}$ and $\ket{\veck_b}$ depend parametrically on $\vecq$.
Operator $\hat{A}_1$ antisymmetrizes the product wavefunction $\ket{X_b}
\ket{\veck_b}$ with respect to the coordinates of the continuum
electron\cite{Spanner09a,Spanner13a}: 
\begin{align}
  \hat{A}_1 & = \frac{1}{\sqrt{N}} \left[ \hat{I} - \sum_{i=1}^{N-1} \hat{P}_{iN} \right], \label{eqn:antisymmetrizer}
\end{align}
where $\hat{I}$ is the $N$-electron identity operator, and operator
$\hat{P}_{iN}$ permutes coordinates of the $i$-th and $N$-th electrons.
Similar to the neutral-space identity $\hat{\bfone}_0$, 
the nuclear wavefunction in $\hat{\bfone}_1$ is expanded in
terms of the harmonic vibrational states $\ket{\bfm}$ of the reference
potential $v_{\rm ref}\left(\vecq\right)$. Because we are not interested in
processes involving multiple ionization, the sum in eq.~\ref{eqn:identity} is
truncated after $\hat{\bfone}_1$.

Operator $\hat{\bfone}$, as well as the individual $\hat{\bfone}_i$ operators
are idempotent.  Identity operators within different subspaces are assumed to
commute:
\begin{align}
  \left[\hat{\bfone}_0,\hat{\bfone}_1\right] & =0. \label{eqn:identity-commutator}
\end{align}
This condition is equivalent to the strong orthogonality assumption for all
$\ket{\veck_b}$:
\begin{align}
  \braket{\phi^\rmD_{ba}}{\veck_b} & = 0, \label{eqn:strong-orthogonality} \\
  \ket{\phi^\rmD_{ba}} & = \sqrt{N} \braket{X_b}{\Phi_a}, \label{eqn:dyson}
\end{align}
where $\ket{\phi^\rmD_{ba}}$ is the Dyson orbital for the ionization of the
neutral state $\ket{\Phi_a}$, forming cation $\ket{X_b}$.

We now insert operator $\hat{\bfone}$ of the Eq.~\ref{eqn:identity} to the left
and to the right of each propagator and interaction operator in
Eq.~\ref{eqn:dipole}.  Following the usual SFA assumptions\cite{Becker05a,Smirnova13a,Smirnova13b}, we also
replace $\hat{U}_\rmT\left(t,t_1\right)$ by a product of the $(N-1)$-electron
ion propagator $\hat{U}_\rmI\left(t,t_1\right)$ and the Volkov propagator
$\hat{U}_\rmV\left(t,t_1\right)$:
\begin{align}
  \hat{U}_\rmT\left(t,t_1\right) & \approx \hat{U}_\rmI\left(t,t_1\right) 
                                           \hat{U}_\rmV\left(t,t_1\right), \label{eqn:ion-volkov} \\
  \hat{U}_\rmV\left(t,t_1\right) \ket{\veck} & = 
      \ket{\veck - \frac{e}{\hbar} \vecA\left(t_1\right) + \frac{e}{\hbar} \vecA\left(t\right)} \times \nonumber \\
    & \exp\left(\frac{-\I\hbar}{2m} \int_{t_1}^t dt' \left(\veck - \frac{e}{\hbar} \vecA\left(t_1\right) 
                   + \frac{e}{\hbar} \vecA\left(t'\right) \right)^2 \right), \label{eqn:volkov} \\
  \braket{\vecr}{\veck} &= \frac{1}{\left(2\pi\right)^{3/2}} \E^{\I \veck\cdot\vecr}, \label{eqn:volkov-state}
\end{align}
where $\vecA\left(t\right)$ is the vector-potential of the laser field, and
Volkov states $\ket{\veck}$ are taken in the length gauge.  The explicit
coordinate representation of $\ket{\veck}$ normalized to
$\delta\left(\veck-\veck'\right)$ is given by Eq.~\ref{eqn:volkov-state}.  In
the simplest form of the SFA used presently, the continuum state
$\ket{\veck_b}$ does not depend on the nature of the binding potential. For
simplicity, we therefore omit the subscript $b$ in $\ket{\veck_b}$ from now on.
(The dependence of $\ket{\veck}$ on the binding potential can of course be
re-introduced if necessary\cite{Smirnova13a,Smirnova13b,Popruzhenko14a}, without materially affecting the overall
form of the MC-SFA-GWP working expressions.) Finally, propagator $\hat{U}_\rmI$
is defined similar to Eq.~\ref{eqn:propagators}, with the $(N-1)$-electron
vibronic Hamiltonian given by $\hat{H}_\rmI$.

Neglecting contributions due to correlation-driven inelastic electron
scattering in the continuum\cite{Spanner09a,Spanner13a}, we obtain:
\begin{widetext}
\begin{align}
   \vecd\left(t\right) & = -\frac{\I}{\hbar}
      \int_{t_0}^{t} d t_1
      \int d\veck 
      \sum_{a'' \bfn''} 
      \sum_{b' \bfm'}
      \sum_{b \bfm}
      \sum_{a''' \bfn'''}
      C_{a''\bfn''}^*\left(t\right)
      \vecR_{b'\bfm'a''\bfn''}^*\left(\veck - \frac{e}{\hbar} \vecA\left(t_1\right) 
                                            + \frac{e}{\hbar} \vecA\left(t\right) \right) \times \nonumber \\ &
      D_{b'\bfm'b\bfm}\left(t,t_1\right)
      \vecF\!\left(t_1\right) \cdot \vecR_{b\bfm a'''\bfn'''}\left(\veck\right)
      C_{a'''\bfn'''}\left(t_1\right) \E^{-\I\phi_d\left(\veck,t,t_1\right)}
 + \hbox{\rm c.c.}, \label{eqn:dipole:resolved} \\
 \vecR_{b\bfm a\bfn}\left(\veck\right) & = 
     e \bra{\bfm} \left[ \braopket{\veck}{\vecr}{\phi^\rmD_{b a}} 
                       + \braket{\veck}{\vec{\phi}^\rmC_{b a}} \right] \ket{\bfn},
     \label{eqn:ionization:dipole} \\
 \ket{\vec{\phi}^\rmC_{ba}} & = \sqrt{N} \braopket{X_b}{\sum_{i}^{N-1}\vecr_i}{\Phi_a}, \label{eqn:craddle} \\
 C_{a\bfn}\left(t\right) & = \sum_{a'\bfn'} C_{a\bfn a'\bfn'}\left(t,t_0\right) C_{a'\bfn'}\left(t_0\right), \label{eqn:wp:neutral} \\
 C_{a'\bfn'}\left(t_0\right) &= \bra{\bfn'} \braket{\Phi_{a'}}{\Psi\left(\vec{r},\vec{q},t_0\right)}, \label{eqn:wp:initial} \\
 C_{a\bfn a'\bfn'}\left(t,t_0\right) & = \E^{\I E_\rmN \left(t-t_0\right) / \hbar} \E^{\I \epsilon_{\bfn'} t_0/\hbar} 
                                         \E^{-\I \epsilon_{\bfn} t/\hbar}
                                         \bra{\bfn} \bra{\Phi_a} \hat{U}_0\left(t,t_0\right) \ket{\Phi_{a'}}\ket{\bfn'},
                                         \label{eqn:wp-coefficient:neutral} \\
 D_{b'\bfm'b\bfm}\left(t,t_1\right) & = \E^{\I E_\rmI \left(t-t_1\right) / \hbar} \E^{\I \epsilon_{\bfm} t_1/\hbar} 
                                        \E^{-\I \epsilon_{\bfm'} t/\hbar}
                                        \bra{\bfm'} \bra{X_{b'}} \hat{U}_\rmI\left(t,t_1\right) \ket{X_{b}}\ket{\bfm},
                                        \label{eqn:wp-coefficient:cation} \\
 \phi_d\left(\veck,t,t_1\right) & = 
      \frac{\ip}{\hbar}\left(t-t_1\right) + \frac{1}{2m\hbar} \int_{t_1}^t dt' 
            \left(\hbar\veck - e\vecA\left(t_1\right) + e\vecA\left(t'\right) \right)^2,
      \label{eqn:dipole:phase}
\end{align}
\end{widetext}
\begin{align}
 \ip & = E_\rmI-E_\rmN. \label{eqn:ip}
\end{align}

In eq.~\ref{eqn:ionization:dipole}, $\vecR_{b\bfm a\bfn}\left(\veck\right)$ is
the dipole matrix element for a bound to continuum vibronic transition.  The
first term in eq.~\ref{eqn:ionization:dipole}
($\braopket{\veck}{\vecr}{\phi^\rmD_{b a}}$) arises due to the direct
interaction between laser field and the active electron.  The second
contribution ($\braket{\veck}{\vec{\phi}^\rmC_{b a}}$) is due to
exchange-mediated interaction of the laser field with the inactive core
electrons\cite{Martin76a,Patchkovskii07a,Patchkovskii07b}.  It is described by
the 3-component ``cradle'' vector orbital of eq.~\ref{eqn:craddle} (after
Newton's cradle, where a force acting on one ball in a multi-ball pendulum
causes a different ball to swing\cite{Spanner09a}).  Coefficients
$C_{a\bfn}\left(t\right)$ describe the vibronic wave packet on the neutral surface
at time $t$.  Coefficients $C_{a\bfn a'\bfn'}\left(t,t_0\right)$ and
$D_{b'\bfm'b\bfm}\left(t,t_1\right)$ are respectively the matrix elements of
the neutral and ionic propagators $\hat{U}_0\left(t,t_0\right)$ and
$\hat{U}_\rmI\left(t,t_1\right)$, where $\epsilon_\bfn$ is the energy of
harmonic vibrational state $\ket{\bfn}$.  The explicit form of the
arbitrary-order Taylor expansion of these matrix elements was given in
Ref.~\cite{Patchkovskii14a} and need not be repeated here.  Equivalently, these
matrix elements can also be obtained with MCTDH time
propagation\cite{Mondal14a,Mondal15a,Rao15a,Rao15b,Rao15c,Arnold17a}.  The
phase factors containing $\epsilon_\bfn$ in
eqs.~\ref{eqn:wp-coefficient:neutral}--\ref{eqn:wp-coefficient:cation} arise to
compensate for the presence of the vibrational phase factor in the definition
of the wavefunction Ansatz in ref.~\cite{Patchkovskii14a} (cf. eq. 2 of
Ref.~\cite{Patchkovskii14a}).  In eqs.~\ref{eqn:wp-coefficient:neutral} and
\ref{eqn:wp-coefficient:cation}, we have extracted the rapidly-oscillating
overall phase of the matrix elements, using $E_\rmN$ and $E_\rmI$ as the
characteristic energy of the neutral and cationic manifolds, respectively.
Finally, $\phi_d$ (Eq.~\ref{eqn:dipole:phase}) is the rapidly-varying part of
the HHG dipole phase accumulated from $t_1$ to $t$. The first term is due to
the bound-state dynamics in the cation, with $\ip$ being the characteristic
ionization potential; the second contribution is the Volkov phase of the
continuum electron.

Applying the usual stationary-phase approximation\cite{Smirnova13a,Smirnova13b} for the $\veck$
integral in eq.~\ref{eqn:dipole:resolved}, we obtain:
\begin{widetext}
\begin{align}
   \vecd\left(t\right) & = 
      \int_{t_0} d t_1
      \sum_{a'' \bfn''} 
      \sum_{b' \bfm'}
      \sum_{b \bfm}
      \sum_{a''' \bfn'''}
      \frac{1}{\hbar}
      \left(\frac{\I 2 \pi m}{\hbar\left(t-t_1\right)}\right)^{3/2}
      C_{a''\bfn''}^*\left(t\right)
      \vecR_{b'\bfm'a''\bfn''}^*\left(\veck_\rms - \frac{e}{\hbar} \vecA\left(t_1\right) 
           + \frac{e}{\hbar} \vecA\left(t\right) \right) \times \nonumber \\ &
      D_{b'\bfm'b\bfm}\left(t,t_1\right)
      \vecF\!\left(t_1\right) \cdot \vecR_{b\bfm a'''\bfn'''}\left(\veck_\rms\right)
      C_{a'''\bfn'''}\left(t_1\right) \E^{-\I\phi_d\left(\veck_\rms,t,t_1\right)}
 + \hbox{\rm c.c.}, \label{eqn:dipole:int-k}
\end{align}
\end{widetext}
\begin{align}
 \veck_\rms\left(t,t_1\right) & = - \frac{e}{\hbar\left(t-t_1\right)} \int_{t_1}^t \vecA\left(t'\right) d t' 
      + \frac{e}{\hbar} \vecA\left(t_1\right),  \label{eqn:stationaryk}
\end{align}
where $\veck_\rms$ is the stationary electron momentum at the time of
ionization $t_1$.

Direct application of the stationary-phase approximation to the $d t_1$
integral in eq.~\ref{eqn:dipole:int-k} leads to a generally complex stationary
ionization time $t_\rms$, causing difficulties when the matrix elements of
eq.~\ref{eqn:ionization:dipole} are only known numerically, on the the real
axis. An alternative, real treatment is based on an observation that for all
but the smallest $\Delta t = t-t_1$ values, the phase $\phi_d$
(Eq.~\ref{eqn:dipole:phase}) is dominated by the Volkov phase\cite{Ivanov96a,Smirnova13a,Smirnova13b}. If
the term containing the $\ip$ can be neglected and the field is linearly
polarized, the stationary ionization time $t_\rms$ is then determined by the
condition $\veck_\rms\left(t,t_\rms\right)=0$.  Small, non-zero $\ip$ values
can then be accommodated using a Taylor expansion around $\veck_\rms=0$
point\cite{Ivanov96a}. For elliptically polarized fields, $\veck_\rms=0$ can not
satisfy eq.~\ref{eqn:stationaryk}, as a small transverse initial momentum is
required to bring the stationary trajectory back to the origin. The
general condition determining the stationary ionization time then becomes:

\begin{align}
  \veck_\rms \cdot \frac{\partial \vecA\left(t\right)}{\partial t_\rms} \equiv 
        - \veck_\rms \cdot \vecF\!\left(t_\rms\right) = 0. \label{eqn:stationaryts}
\end{align}
Eq.~\ref{eqn:stationaryts} together with eq.~\ref{eqn:stationaryk} define the
real ionization time $t_\rms$.

An additional complication arises when the vibronic wavefunction is even with
respect to the origin and the laser field is linearly polarized. Then, the matrix
element $\vecR_{b\bfm a'''\bfn'''}\left(\veck_\rms\right)$ in
eq.~\ref{eqn:dipole:int-k} vanishes at $\veck_\rms=0$, and the next order in
the power series determines the overall integral over $d t_1$.  Expanding the
ionization dipole through the first order around $\veck_\rms$ and assuming that
the remaining matrix elements in Eq.~\ref{eqn:dipole:int-k} vary slowly with
$\veck_\rms$, we obtain (also see Appendices \ref{sec:appendix-A}
and \ref{sec:appendix-B}):
\begin{widetext}
\begin{align}
   \vecd\left(t\right) & = 
      \sum_{t_\rms}
      \sum_{a'' \bfn''} 
      \sum_{b' \bfm'}
      \sum_{b \bfm}
      \sum_{a''' \bfn'''}
      \left(\frac{\I 2 \pi m}{\hbar\left(t-t_\rms\right)}\right)^{3/2}
      C_{a''\bfn''}^*\left(t\right)
      \vecR_{b'\bfm'a''\bfn''}^*\left(\veck_\rms - \frac{e}{\hbar} \vecA\left(t_\rms\right) 
          + \frac{e}{\hbar} \vecA\left(t\right) \right) \times \nonumber \\ &
      D_{b'\bfm'b\bfm}\left(t,t_\rms\right)
      \Upsilon_{b\bfm a'''\bfn'''}\left(\veck_\rms,t_\rms\right)
      C_{a'''\bfn'''}\left(t_\rms\right) \E^{-\I\phi_d\left(\veck_\rms,t,t_\rms\right)}
 + \hbox{\rm c.c.}, \label{eqn:dipole:final} \\
 \Upsilon_{b\bfm a'''\bfn'''} & = 
     \vecF\!\left(t_\rms\right) \cdot \vecR_{b\bfm a'''\bfn'''}\left(\veck_\rms\right) 
     2 \pi  \left(\frac{2 m}{e^2 \hbar^2 \vecF^2\!\left(t_\rms\right)}\right)^{1/3} \Ai\left(\zeta\right) \nonumber \\
    & - 2\I\pi  \frac{\partial}{\partial \veck_\rms} \left[ \vecF\!\left(t_\rms\right) 
        \cdot \vecR_{b\bfm a'''\bfn'''}\left(\veck_\rms\right) \right]
        \cdot \left( \frac{\hbar\veck_\rms}{t-t_\rms} - e \vecF\!\left(t_\rms\right) \right)
              \left(\frac{2 m}{e^2 \hbar^2\vecF^2\!\left(t_\rms\right)}\right)^{2/3} \Ai'\left(\zeta\right),
    \label{eqn:ionization:final}
\end{align}
\end{widetext}
\begin{align}
 \zeta & = \left(\frac{2m}{e^2\hbar^2\vecF^2\!\left(t_\rms\right)}\right)^{1/3}\left(E_{b\bfm} - E_{a'''\bfn'''} 
         + \frac{\hbar^2\veck_\rms^2}{2m}\right), \label{eqn:ionization:argument} \\
 E_{b\bfm} &= \bra{\bfm}\braopket{X_b}{\hat{H}_\rmI}{X_b}\ket{\bfm}, \label{eqn:ion-state-energy} \\
 E_{a'''\bfn''} &= \bra{\bfn'''}\braopket{\Phi_{a'''}}{\hat{H}_0}{\Phi_{a'''}}\ket{\bfn'''}, \label{eqn:neutral-state-energy}
\end{align}
where $E_{a'''\bfn''}$ and $E_{b\bfm}$ are respectively energies of the neutral
and cationic vibronic basis functions.  In Eq.~\ref{eqn:ionization:final} we
neglected the explicit time dependence of $\vecF$, which is in the same order
as other terms omitted in deriving eq.~\ref{eqn:ionization:final} (See
Appendix~\ref{sec:appendix-A}). We have also neglected the $\veck_\rms$
dependence of the recombination matrix elements $\vecR_{b'\bfm'a''\bfn''}^*$.
If necessary (for example, in the vicinity of a Cooper minimum), this
dependence can be reintroduced in a similar manner.

Eq.~\ref{eqn:dipole:final} is our final working equation. A closely-related
special-case result was previously obtained by Chiril\v{a} and
Lein\cite{Chirila08a}, and by Madsen et al.\cite{Madsen10a}, which however did
not consider the possibility of state crossings and non-adiabatic vibronic
coupling. The special-case treatment of $\rm SF_6$ by Walters et
al\cite{Walters09a} is similar in spirit, and in principle includes all effects
considered here.  Another closely-related result by Faisal\cite{Faisal10a}
includes rotational motion (neglected here), but restricts
vibrational dynamics to a single surface. In contrast, the treatment of
Ref.~\cite{Le12a}, while superficially similar, completely neglects all
sub-cycle effects arising due to the motion on the intermediate cationic
surfaces. These effects are also completely neglected by
Ref.~\cite{Lisinetskaya11a}, which however treats the vibronic dynamics in the
neutral manifold. Finally, treatments of Patchkovskii and
Schuurman~\cite{Patchkovskii14a}, Varandas et
al\cite{Mondal14a,Mondal15a,Rao15a,Rao15b,Rao15c}, and Arnold et
al\cite{Arnold17a} fully treat cationic vibronic dynamics, but neglect neutral
dynamics and their coupling to the HHG process.

\section{Computational details\label{sec:details}}

\subsection{Electronic structure calculations\label{sec:details:qchem}}

All electronic structure calculations used correlation-consistent, valence
triple-$\zeta$ basis set, augmented with diffuse functions
(``aug-cc-pVTZ''\cite{Dunning89a,Kendall92a}).  Calculations of the neutral $\rm CH_4$ geometry
and vibrational Hessian used second-order M{\o}ller-Plesset (MP2) correlation
treatment, with the $1s$ electrons on the carbon atom left uncorrelated.
Optimized C--H bond length ($r_0=1.110$\AA) and unscaled harmonic vibrational
frequencies ($\nu_1 ... \nu_4 = $ $3069$, $1589$, $3204$, $1356$~cm$^{-1}$) are
in a good agreement with the experimental equilibrium geometry and fundamental
frequencies ($r_e=1.094$\AA, $\nu_1 ... \nu_4 = $ $2916$, $1534$, $3019$,
$1306$~cm$^{-1}$\cite{Herzberg66a}).

\begin{figure}[thbp]
\includegraphics[width=8.26cm]{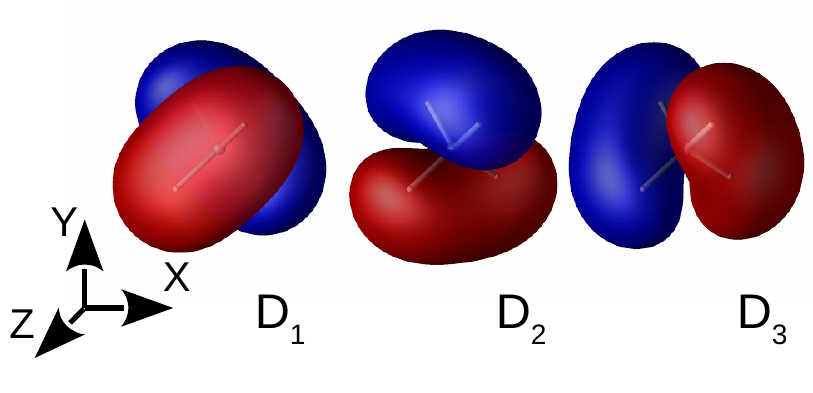}
\caption{\label{fig:dyson} (Color online) MR-CIS Dyson orbitals for the three degenerate
components of the lowest ($^2F_2$) ionization channel in $\rm CH_4$. The
orbitals are shown at the high-symmetry $\rm T_d$ point. Isosurface levels are
$0.02~a_0^{-3/2}$.
}
\end{figure}

The equilibrium neutral geometry corresponds to a symmetry-required triple
degeneracy point of the $^2F_2$ electronic ground state of the methane
cation\cite{Jacox03a}. The electronic structure in the vicinity of this degeneracy
point was explored by displacing all pairs of Cartesian coordinates by
$\pm0.02$\AA, for the total of $450$ unique distorted structures. The reference
geometry was taken in the standard setting of the $T_d$ group, with the $C_2$
axes along the main Cartesian directions (See Fig.~\ref{fig:dyson}).
Symmetry was used neither to reduce the number of the displaced geometries,
nor in the electronic structure calculations at these geometries.

Electronic wavefunctions of the three lowest cation states correlating to the
$^2F_2$ state were calculated using multi-reference configuration interaction
with single excitations treatment (MR-CIS). All configurations within the
minimal complete active space with 7 electrons in 4 frontier orbitals
[CAS(7,4)] were included in the reference set. Single-particle orbitals were
optimized in a CAS(7,4) self-consistent field (CASSCF) calculation, with the
energies of the three cation states correlating to $^2F_2$ weighted equally.
The neutral wavefunctions were determined within the CIS space, using the
lowest closed-shell determinant constructed from the cation-optimized orbitals
as the reference. The MR-CIS vertical ionization potential (MR-CIS: $13.14$~eV)
slightly underestimates the experimental value (Expt:
$13.60$~eV\cite{Lias17a,Berkowitz15a}).  Including double excitations in the CI expansion
(MR-CISD; Langhoff-Davidson +Q correction\cite{Langhoff74a} not included) leads
to an overestimation of the vertical IP by a similar amount (MR-CISD:
$13.97$~eV). Given the large number of the distorted geometries, further
increase in the CI size is not practical.  It is in principle possible to
adjust the calculated vertical IP to match the experimental value. However, the
experimental vertical ionization potential is somewhat uncertain, with a very
very broad, complex peak in the $13.2$--$14.0$~eV range\cite{Kimura81a,Berkowitz15a}.
Because all qualitative features of the energy surfaces are already adequately
described at the MR-CIS level, all subsequent calculations use MR-CIS
wavefunctions and energies. Due to the underestimation of the vertical IP, we
expect ionization rates (Eq.~\ref{eqn:ionization:final}) to be somewhat
overestimated.

\subsection{Diabatization procedure\label{sec:details:diabat}}

Evaluation of matrix elements entering Eq.~\ref{eqn:dipole:final} requires that
both the potential energy surfaces and and the wavefunctions are smooth and
continuous in the vicinity of the expansion point.  It is therefore necessary
to diabatize the MR-CIS states. Construction of quasi-diabatic states is a
long-standing, and still active area of research (See Ref.~\cite{Yarkony12a}
for an overview), with many competing prescriptions available in the
literature.  An appealing, generally-applicable approach for constructing local
quasi-diabatic surfaces involves fitting adiabatic energy gradients and
derivative couplings at selected points in the nuclear coordinate
space\cite{Schuurman07a,Papas08a,Dillon11a}.  Unfortunately, this technique
does not offer an easy access to the transformed quasi-diabatic wavefunctions,
which are needed for evaluating the somewhat non-standard matrix elements
$\Upsilon$ and $\vec{R}$
(Eqs.~\ref{eqn:ionization:dipole},\ref{eqn:ionization:final}). Here, we retain
the quadratic vibronic Hamiltonian form of Ref.~\cite{Schuurman07a}, but revert
to the older maximum-overlap approach\cite{Gadea90a,Petsalakis91a,Pacher91a} to
directly determine the diabatic wavefunctions:
\begin{align}
  \Psi^{\rm dia}_{ja} & = \sum_i U_{jia} \Psi^{\rm ad}_{ia}, & \label{eqn:diabatic:transform} 
\end{align}
where $\Psi^{\rm dia}_{ja}$ is the $j$-th quasi-diabatic state at geometry $a$,
$\Psi^{\rm ad}_{ia}$ is the $i$-th adiabatic state at the same geometry.  The
unitary transformation $U_{jia}$ connects the two wavefunction spaces. At
each displaced geometry $(a)$, the optimal matrix $\bfU_a$ is
determined\cite{Gadea90a,Petsalakis91a,Pacher91a,Golebiewski61a,Murrell60a,Cederbaum89a}
by the overlap matrix $\bfS_a$ between the adiabatic wavefunctions at the
displaced geometry and reference wavefunctions $\Psi^{\rm ref}_{k}$ (see
Fig.~\ref{fig:dyson}):
\begin{align}
  S_{ika} & = \braket{\Psi^{\rm ad}_{ia}}{\Psi^{\rm ref}_{k}}, & \label{eqn:diabatic:overlap} \\
  \bfU_a^* & = \left(\bfS_a^\dagger \bfS_a\right)^{-1/2} \bfS_a^{\dagger}. & \label{eqn:diabatic:umat}
\end{align}
For methane, reference wavefunctions $\Psi^{\rm ref}$ are taken at the high-symmetry $T_d$
geometry, treated within the $D_2$ sub-group. Neutral manifold uses the $^1A_1$ ground-state 
wavefunction, while cationic manifold is referenced to the lowest $^2B_1$, $^2B_2$, and $^2B_3$
wavefunctions.

Once the diabatization transformation $\bfU_a$ is determined, the quadratic
vibronic Hamiltonian is obtained with a least-squares fit to the diabatic
Hamiltonian matrices at the displaced geometries. Electronic matrix elements
are calculated directly from the transformed diabatic wavefunctions
(eq.~\ref{eqn:diabatic:transform}) as described elsewhere\cite{Lowdin55a,Patchkovskii07b} (also see
Appendix~\ref{sec:appendix-C}). For the methane molecule in the vicinity of the
equilibrium neutral geometry, the resulting vibronic Hamiltonian coincides with
the result of \cite{Schuurman07a}, while the residual derivative couplings are
found to be numerically negligible. Direct non-linear minimization of the
quadratic vibronic Hamiltonian fit with respect to the diabatization parameters
$\bfU_a$ leaves the result unchanged, confirming that the transformation of
Eq.~\ref{eqn:diabatic:umat} represents a local optimum.

It should be noted that the quadratic vibronic Hamiltonian of the $^2F_2$ manifold
of $\rm CH_4^+$ determined presently focuses on the small region of 
configuration space in the vicinity of the neutral equilibrium geometry.
It is unbound from below, and corresponds to a dissociative state. As a result,
it is not capable of describing long-time dynamics of the cation, and should
not be compared globally to the more conventional, spectroscopic surfaces,
such as used in Refs.~\cite{Mondal14a,Mondal15a}.

\subsection{Dyson orbitals and dipole matrix elements\label{sec:details:prop}}

Dyson orbitals (Eq.~\ref{eqn:dyson}) corresponding to the ground-state
($^2F_2$) ionization channel of the $\rm CH_4$ molecule are shown in
Fig.~\ref{fig:dyson}. All three components are strongly 1-electron allowed,
with the Dyson orbital norm of $0.916$. Visually, these Dyson orbitals are
indistinguishable from the triply-degenerate HOMO of the neutral species.  The
dominant ``cradle'' terms (Eq.~\ref{eqn:craddle}; not shown) are derived from
the ($s$-like) $2a_1$ molecular orbital of the methane molecule. 

\begin{figure}[thbp]
\includegraphics[width=8.26cm]{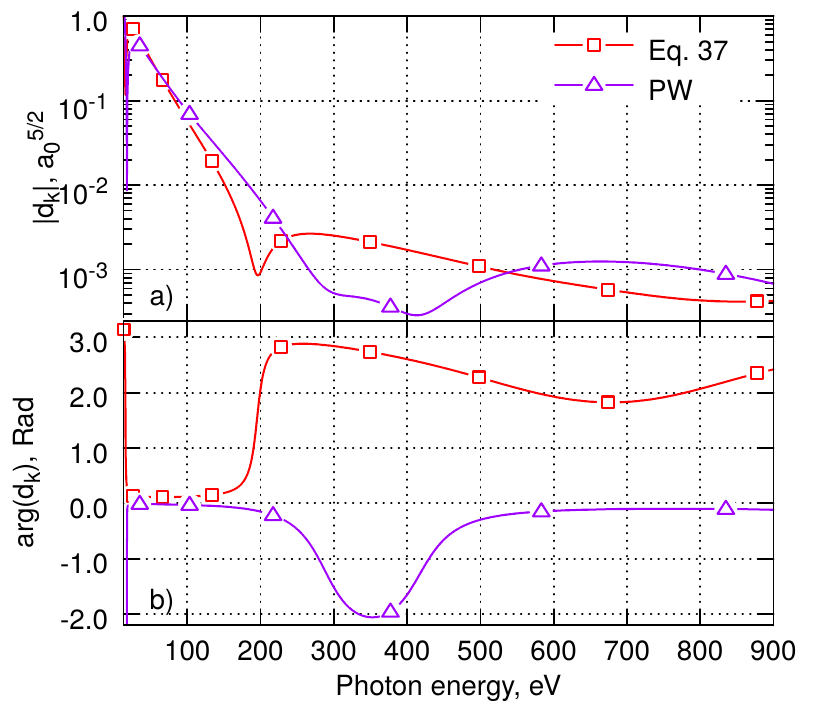}
\caption{\label{fig:cross} (Color online) Magnitude (panel a) and phase (panel b) of the
dipole photoionization matrix elements (Eq.~\ref{eqn:dipole:electronic}) for
the $D_3$ component of the $^2F_2$ ionization channel at the high-symmetry
point (Fig.~\ref{fig:dyson}). The $\vec{k}$ vector is taken along the positive
$X$ Cartesian direction.  The ``cradle'' contributions (Eq.~\ref{eqn:craddle})
are significant beyond $\approx 150$~eV.  Omitting the ``cradle'' terms leads
to the ``planewave'' (PW) curve.
}
\end{figure}

It is instructive to examine the electronic part of the photoionization matrix
elements:
\begin{align}
 \vec{d}_e & = e \left[ \braopket{\veck}{\vecr}{\phi^\rmD_{b a}} + \braket{\veck}{\vec{\phi}^\rmC_{b a}} \right] 
               & \label{eqn:dipole:electronic}
\end{align}
(see Eq.~\ref{eqn:ionization:dipole}) a little closer. Due to the high symmetry
of the $\rm CH_4$ molecule, it is sufficient to constrain the photoelectron
observation direction to one of the Cartesian axes (See
Figs.~\ref{fig:dyson} and \ref{fig:cross}). The dominant first term in the matrix
element is identical to the first Born approximation in photoionization, known
to be qualitatively inaccurate close to the ionization threshold. Beyond
$30$~eV photon energy, however, the calculated cross-sections are in 
a (fortuitously) good agreement with experimental
data\cite{Lee77a,Samson89a,Au93a,Au93b,Kameta02a,Berkowitz15a} and accurate
calculations\cite{Reinsch85a,Cacelli88a,Tenorio16a}. For example,
Eq.~\ref{eqn:dipole:electronic} yields photoionization cross-section of
$\approx17$~Mbarn at $30$~eV photon energy (expt:
$\approx14$Mbarn\cite{Tenorio16a}), decreasing to $\approx5$~Mbarn at $60$~eV
(expt: $\approx2$~Mbarn\cite{Tenorio16a}), and to $\approx1.2$~Mbarn at $90$~eV
(expt: $\approx0.6$~Mbarn\cite{Tenorio16a}). 

The ``cradle'' terms become important\cite{Martin76a} beyond $\approx150$~eV
photon energy (See Fig.~\ref{fig:cross}), leading to an $\approx0.4$~kbarn
minimum in the calculated cross-sections at $\approx 196$~eV. It is followed by
a maximum ($\approx7$~kbarn) at $295$~eV, (again fortuitously) close to where
the cross-section is expected to increase due to the K-edge intensity
borrowing\cite{Amusia75a}. Unfortunately, we are not aware of reliable
final-state resolved calculations or measurements in methane for photon
energies above $100$~eV, where ionization from the inner $2a_1$ and $1a_1$
shells dominates the overall cross-section\cite{Berkowitz15a}. It is therefore
unclear whether the photoionization matrix elements in Fig.~\ref{fig:cross}
provide a reasonable description of methane photoionization beyond $90$~eV.
Nonetheless, the minimum and the associated $\pi$ phase jump at $196$~eV serve
to illustrate an important point (see below), which remains qualitatively valid
even it occurs at a different energy in the actual $\rm CH_4$ molecule than in
our crude calculations here.

\subsection{Nuclear autocorrelation and MC-SFA-GWP HHG calculations\label{sec:details:ac}}

Nuclear autocorrelation functions are a special case of the vibronic matrix
elements $D_{b'\bfm'b\bfm}\left(t,0\right)$ of
Eq.~\ref{eqn:wp-coefficient:cation}, with $b'=b$ and $m'=m$. Both
$D_{b'\bfm'b\bfm}$ and $C_{a'''\bfn'''}$ matrix elements are propagated in time
numerically, using the 4-th order Runge-Kutta integrator with a uniform time
step\cite{NR03}.  Calculations of the autocorrelation function used time step of
$0.02$~au[t] ($\approx0.48$~as).  Short-time autocorrelation functions
calculated presently numerically coincide with the analytical
results~\cite{Patchkovskii14a}, for all times where the power-series expansion
in Ref.~\cite{Patchkovskii14a} converges.

Calculations of the time-dependent MC-SFA-GWP dipole $\vecd\left(t\right)$
(Eq.~\ref{eqn:dipole:final}) used a time step corresponding to $1/3$ of the
Nyquist limit for the cut-off harmonics.  The resulting time step ranged from
$0.386$~au[t] ($\approx 9.3$~as) for the $800$~nm, 300~TW~cm$^{-2}$ driving
field, to $0.037$~au[t] ($\approx 0.89$~as) for the $1.6\mu$m, 1~PW~cm$^{-2}$
driver. We did verify that all calculations are converged with respect to the
time step.  The high-harmonics spectrum was calculated from the Fourier
transform of the time-dependent dipole\cite{Smirnova13a,Smirnova13b}.  No window function was
applied.

The vibronic wavefunction expansion was performed in the basis of multi-dimensional
harmonic oscillator functions of the neutral species. Rotational and
translational modes were excluded. Calculations of the auto-correlation
functions allowed up to $12$ quanta in any of the remaining $9$ normal modes.
This choice is fully converged with respect to the basis set up to at least
$4.5$~fs evolution time, but is too computationally expensive for MC-SFA-GWP
calculations, where the vibronic matrix elements need to be recalculated for
each trajectory. Instead, MC-SFA-GWP calculations limited the vibronic basis to
at most $4$ quanta in any of the vibrational modes. This choice changes the
autocorrelation function by less than $0.2$\% at $1.7$~fs delay ($800$-nm
cut-off trajectory), increasing to $3$\% at $2.6$~fs ($1200$-nm cut-off) and
$25$\% at $3.4$~fs ($1600$-nm cut-off). Our MC-SFA-GWP results are therefore
sufficiently converged at $800$ and $1200$~nm, but are only qualitative for the
high-energy part of the spectrum at $1800$~nm (beyond $160$/$500$~eV at
$300$/$1000$~TW~cm$^{-2}$).

All MC-SFA-GWP calculations were performed for laser electric field polarized
along the molecular $X$ axis.  Due to the high symmetry of the $\rm CH_4$
molecule, we expect orientational averaging to play only a minor role for this
system. Only the short trajectories were considered.

\section{Results and discussion\label{sec:results}}

\subsection{Free-cation dynamics and autocorrelation function\label{sec:results:auto}}

\begin{figure}[thbp]
\includegraphics[width=8.26cm]{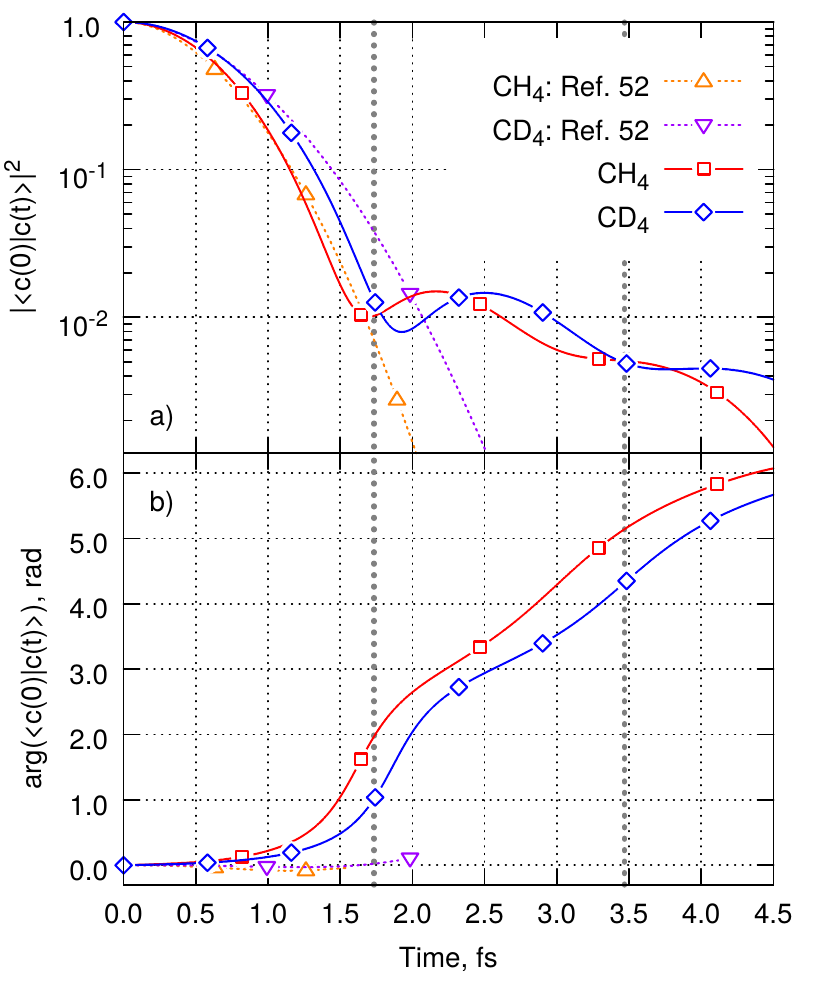}
\caption{\label{fig:auto} (Color online) Nuclear autocorrelation function in methane cation.
Initial nuclear wave packet $\ket{c(0)}$ is the ground vibrational state of
the neutral species. Solid red line: $\rm CH_4$, this work.  Solid blue line:
$\rm CD_4$, this work. Dotted orange line: $\rm CH_4$ using field-diabatized
cationic energy surface (Ref.~\cite{Patchkovskii09a}). Dotted purple line: $\rm
CD_4$ from Ref.~\cite{Patchkovskii09a}.  Panel (a) shows the squared absolute
magnitude of the autocorrelation function.  Panel (b) gives the phase of the
autocorrelation function. Linear phase due to the zero-point energy has been
subtracted.  Dotted, gray vertical lines indicate time delays for the cut-off
trajectories for an $800$ and $1600$-nm driving fields.
}
\end{figure}

The simplest summary of the subcycle nuclear dynamics is provided by the
nuclear autocorrelation function, shown in Fig.~\ref{fig:auto} for both $\rm
CH_4$ and $\rm CD_4$ cations, using neutral vibrational ground state as the
initial wavefunction at time zero. For comparison, we show autocorrelation
functions for the same species and initial conditions from
Ref.~\cite{Patchkovskii09a}.  At very short times (up to $\approx 1.2$~fs), all
four autocorrelation functions are Gaussian in time (please note the linear-log
scale of the Fig.~\ref{fig:auto}a), which is a generic feature of short-time
autocorrelation functions\cite{Patchkovskii09a,Tannor07a}.  The agreement
between the data sets is remarkably good, considering the differences in the
theoretical treatment. In Ref.~\cite{Patchkovskii09a}, the autocorrelation
functions are calculated on a single component of the $^2F_2$ electronic
surface, with the degeneracy lifted by applying an intense ($\approx
2.6$~V\AA$^{-1}$) static electric field. Because the components of the $^2F_2$
state are coupled by the electric field (MR-CIS transition dipole of $\approx
0.7$~Debye), the surface shape is also modified, allowing the single-surface
dynamics to mimic the more complex vibronic dynamics at early times. 

At the same time, the single-surface treatment does not allow for the
``orbiting'' motion of the wave packet around the conical intersection
point, precluding the possibility of revivals on a few-femtosecond
time scale.  Beyond $\approx1.2$~fs, a single, field-diabatized surface no
longer adequately represents vibronic dynamics in this system. For $\rm CH_4$
($\rm CD_4$), the population of the initial surface reaches a minimum of $52$\%
($51$\%) at $1.5$~fs ($1.7$~fs), then begins to increase again as a fraction of
the wave packet completes a half-revolution around the CI. Correspondingly, the
autocorrelation function changes sign (See Fig.~\ref{fig:auto}b) and undergoes
a half-revival at $2.15$~fs ($2.50$~fs). For both $\rm CH_4$ and $\rm CD_4$,
the nuclear autocorrelation factor in HHG reaches $1.5$\% at the half-revival
-- more than an order of magnitude above the expected damping factor due to the
Gaussian decay\cite{Patchkovskii09a}.  A full wave packet revival appears as a
shoulder at the $0.5$\% level, at $\approx3.6$~fs ($\approx 4.2$~fs).
Similar, non-Gaussian intermediate-time decay and unexpected revivals has been previously
predicted in other non-adiabatic systems\cite{Forster13a,Patchkovskii14a,Arnold17a}.

The short-time autocorrelation functions in Fig.~\ref{fig:auto} are
qualitatively consistent with the previously reported calculations by Mondal
and Varandas~\cite{Mondal14a,Mondal15a}. These authors find the first
half-revival at $\approx 2.4$~fs for $\rm CH_4^+$ ($\approx 2.8$~fs for $\rm
CD_4^+$). The full revivals are also found at times comparable to our results.
At the same time, the maximum of the $\rm CD_4$/$\rm CH_4$ autocorrelation
functions ratio calculated in Ref.~\cite{Mondal15a} is much higher ($\approx 6$,
compared to $\approx 2.7$ here.  It is also reached at a later time ($\approx
1.85$~fs, compared to $\approx 1.5$~fs here). Both discrepancies are well
within the range of available experimental resolution, and would be interesting
to explore.

Furthermore, in contrast to Ref.~\cite{Mondal14a,Mondal15a}, we cannot
attribute the revivals to the dynamics reaching the Jahn-Teller minima on the
cationic energy surface.  Calculated expectation values of normal coordinates
on each electronic surface evolve monotonically at least until $3.5$~fs. The
Cartesian displacements in the centre of mass coordinate system remain small
(less than $0.02$~\AA), with the structure remaining at the $D_{2d}$ symmetry,
indicating that the minimum is not reached at the times relevant for the PACER
experiments at mid-IR wavelengths.

\begin{figure}[thbp]
\includegraphics[width=8.26cm]{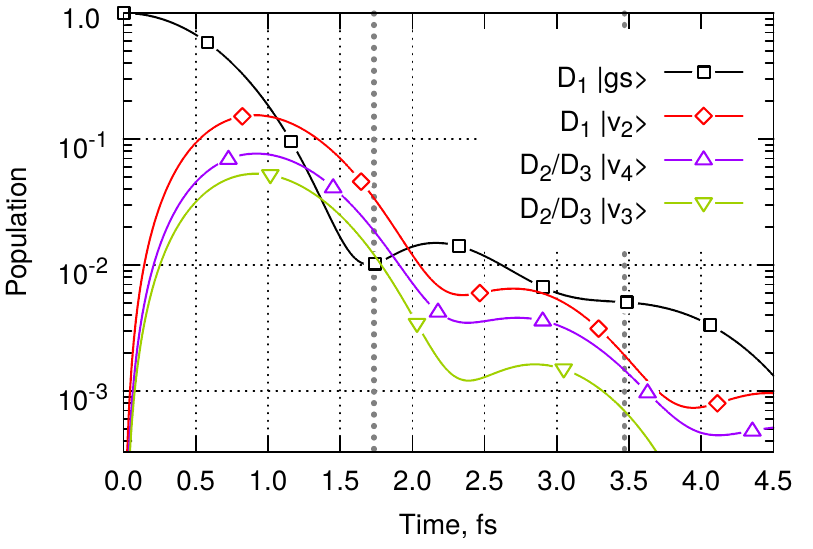}
\caption{\label{fig:autopop} (Color online) Transient population of dominant singly-excited
vibronic basis functions in $\rm CH_4^+$.  The initially-prepared wave packet is
ground vibrational state of the neutral molecule, placed on the $D_1$ diabatic
surface ($D_1 \ket{{\rm gs}}$: black squares). This vibronic basis function is
directly coupled to the singly-excited $\nu_2$ mode on the same electronic
surface ($D_1 \ket{\nu_2}$: red diamonds), singly-exited $\nu_4$ mode on the
$D_2$ and $D_3$ surfaces ($D_2/D_3 \ket{\nu_4}$: purple triangles), and
singly-excited $\nu_3$ mode on the same surfaces ($D_2/D_3 \ket{\nu_3}$: green
inverted triangles).
}
\end{figure}

It is therefore instructive to examine the origin of the oscillations in the short-time
autocorrelation function in a little more detail. The square modulus of the
autocorrelation function measures the population of the initially-populated
vibronic basis function ($D_1 \ket{{\rm gs}}$ in Fig.~\ref{fig:autopop}). The
strongest vibrational coupling of $\ket{{\rm gs}}$ on the \textit{same}
quasi-diabatic electronic surface is to a vibrational basis function with a
single quantum in the $\nu_2$ vibrational mode ($\ket{\nu_2}$: $e$ symmetry,
$1589$~cm$^{-1}$, $\rm H-C-H$ bending motion). The strongest coupling to the
other two quasi-diabatic surfaces ($D_2$ and $D_3$) involve single excitation
of the $\nu_4$ ($\ket{\nu_4}$: $t_2$, $1356$~cm$^{-1}$, $\rm H-C-H$ bend) and
$\nu_3$ ($\ket{\nu_3}$: $t_2$, $3204$~cm$^{-1}$, $\rm C-H$ stretch) normal
modes. All of these couplings are of a similar magnitude.  The number of
significant couplings increases rapidly with the vibrational excitation level;
once two or more vibrational quanta have been excited, the probability of the
wave packet returning to the initial state $D_1\ket{{\rm{gs}}}$ becomes
negligible.

The short-time evolution of the population of the initially-populated and
singly-excited vibronic basis functions in $\rm CH_4^+$ is illustrated in
Fig.~\ref{fig:autopop}.  One can immediately see that the populations of all
five dominant singly-excited functions ($D_1\ket{\nu_2}$, $D_2/D_3\ket{\nu_4}$, and
$D_2/D_3\ket{\nu_3}$) oscillate in-phase at least up to $4.5$~fs. The
oscillations are out-of-phase with the autocorrelation magnitude
($D_1\ket{{\rm{gs}}}$ population), indicating the back-and-forth population
transfer. This behavior is easy to rationalize by considering a
reduced-dimensionality model, consisting of just the six essential states
listed above. The corresponding model Hamiltonian is:
\begin{align}
  H_{\rm mod, 1} & = {\renewcommand{\arraystretch}{1.0}\left(\begin{array}{llllll}
                     0   & b_1 & b_2           & b_2 & b_3           & b_3 \\
                     b_1 & d_1 &               &     &               &     \\
                     b_2 &     & d_2           &     & \mbox{\huge0} &     \\
                     b_2 &     &               & d_2 &               &     \\
                     b_3 &     & \mbox{\huge0} &     & d_3           &     \\
                     b_3 &     &               &     &               & d_3 \\
                     \end{array}\right)}, & \label{eqn:hmod:1}
\end{align}
where all elements not in the first row or column, or on the main diagonal, 
are zero.  As long as all $d_i\approx d$, this Hamiltonian is equivalent
to a two-level system:
\begin{align}
  H_{\rm mod, 2} & = {\renewcommand{\arraystretch}{1.0}
                      \left(\begin{array}{ll} 0 & b_0 \\ b_0 & d \\ \end{array}\right)}, & \label{eqn:hmod:2} \\
  b_0 & = \sqrt{b_1^2 + 2b_2^2 + 2b_3^2}, & \label{eqn:hmod:b0}
\end{align}
where the transformed basis is:
\begin{align}
  \phi_1 & = D_1 \ket{{\rm gs}}, \label{eqn:hmod:phi1} \\
  \phi_2 & = b_0^{-1} \big\{ b_1 D_1 \ket{\nu_2} + b_2 D_2 \ket{\nu_4} + b_2 D_3 \ket{\nu_4}  \\
         & \phantom{ = b_0^{-1} \big\{ b_1 D_1 \ket{\nu_2}}
                                                 + b_3 D_2 \ket{\nu_3} + b_3 D_3 \ket{\nu_3}
             \big\}. \label{eqn:hmod:phi2}
\end{align}
The $\phi_2$ basis function corresponds to a vibronic wave packet
component ``orbiting'' the conical intersection. In the nuclear coordinate space,
$\ket{\phi_2}$ forms a prolate spheroidal shell around the conical
intersection.  The oscillatory behavior in the autocorrelation function (ie
the population of the $\ket{\phi_1}$ basis function) is due to the interference
between the two eigenstates of Eq.~\ref{eqn:hmod:2}, while the overall decay of
the signal in this model arises from the imaginary part of $d$ ($\Im{(d)}<0$).
To the order $O(d^2)$, the difference between the eigenvalues of $H_{\rm mod, 2}$
is:
\begin{align}
  \Delta E_{\rm mod, 2} & \approx 2 b_0. & \label{eqn:hmod:de}
\end{align}
Thus, the oscillation frequency of the short-time autocorrelation function is
determined by the strength of the coupling between the initially-prepared
component of the wave packet and its decaying part ``orbiting'' the
conical intersection.

As can be clearly seen from Eq.~\ref{eqn:hmod:b0}, non-adiabatic vibronic
coupling between the electronic surfaces plays a key role in the initial wave
packet decay. If vibronic coupling is neglected, the initial decay is slowed
down dramatically (data not shown), as was seen before for the benzene
cation\cite{Patchkovskii14a}. The resulting lower isotope effects are no longer
compatible with experimental data (See Section~\ref{sec:results:isotope}
below). A similar observation was made in Ref.~\cite{Madsen10a}, which
neglected the vibronic coupling.

\begin{figure}[thbp]
\includegraphics[width=8.26cm]{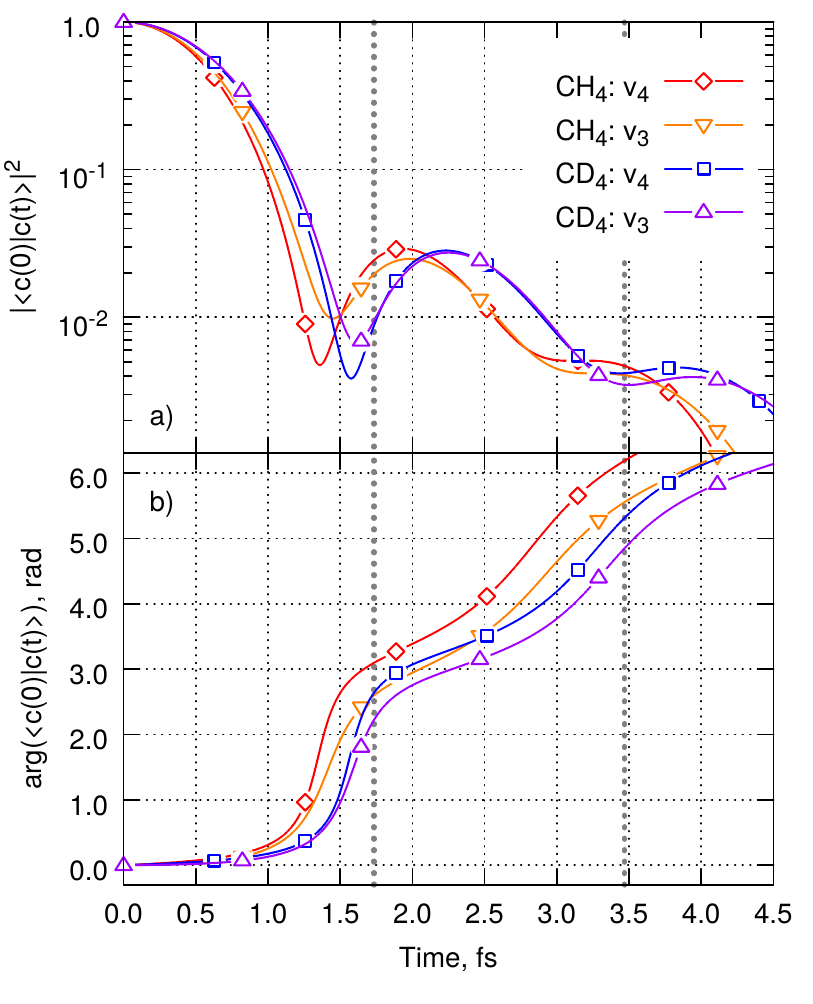}
\caption{\label{fig:auto-vib} (Color online) Nuclear autocorrelation function in vibrationally
excited methane cation.  Initial nuclear wavefunction $\ket{c(0)}$ has a single
vibrational quantum in the asymmetric bend ($\nu_4$, $1356$/$1028$~cm$^{-1}$ in
$\rm CH_4$/$\rm CD_4$) or asymmetric stretch ($\nu_3$, $3204$/$2379$~cm$^{-1}$)
vibrational modes.  Panels and the scale of the plots is the same as in
Fig.~\ref{fig:auto}; see Fig.~\ref{fig:auto} caption for further details.
}
\end{figure}

Coordinate dependence of the tunneling-ionization matrix elements
(eq.~\ref{eqn:ionization:final}) can lead to substantial reshaping of the
neutral vibrational wave packet upon ionization\cite{Goll06a}. In methane,
ionization primarily populates $\nu_4$ (asymmetric bend) and $\nu_3$
(asymmetric stretch) modes. Nuclear autocorrelation functions for
single-quantum excitation of these two modes in $\rm CH_4$ and $\rm CD_4$ are
shown in Fig.~\ref{fig:auto-vib}.  These autocorrelations are qualitatively
similar to the case of the vibrational ground state (Fig.~\ref{fig:auto}).
However, the first half-revival occurs at an earlier time, and has a
substantially higher magnitude ($\rm CH_4$: $2.9$/$2.5$\% at $1.93$/$1.98$~fs
for $\nu_4$/$\nu_3$; $\rm CD_4$: $2.8$/$2.7$\% at $2.23$/$2.25$~fs for
$\nu_4$/$\nu_3$ respectively).  The faster short-time dynamics of vibrationally
excited states has been noted
before\cite{Chirila08a,Chirila09b,Patchkovskii09a,Falge10a,Rao15b}.  This
property will be important for understanding the HHG spectra at longer
wavelengths (Section~\ref{sec:results:hhg} below).

Overall, non-adiabatic nuclear autocorrelation factors suggest that HHG
emission in methane should persist at much longer times than expected
previously. Even at $4.3$~fs, close to the short-trajectory cut-off for a
$2\mu$m IR driver, the autocorrelation factor remains above $0.2$\% ($0.4$\%
for $\rm CD_4$) -- within the possible measurement range. Furthermore, the
non-adiabatic dynamics around the CI can be expected to lead to a rich PACER
signal, including inverse isotope effects between $1.8$ and $2.3$~fs. A more
detailed investigation of the HHG signal therefore appears justified, and is
attended to in Sections~\ref{sec:results:hhg} and \ref{sec:results:isotope}
below.

\subsection{High-harmonics generation\label{sec:results:hhg}}

For calculations of high-harmonics spectra, we consider three driver
wavelengths ($800$~nm, $1.2$~$\mu$m, and $1.6$~$\mu$m) and two peak intensities
($300$~TW~cm$^{-2}$ and $1$~PW~cm$^{-2}$). For each combination of the
wavelength, intensity, and isotopic species, we perform four simulations,
namely:
\begin{itemize}
\item[``F'':] 
In this simulation, the nuclei are ``frozen'', and are not allowed to move
between ionization and recollision.  To this end, the vibronic matrix elements
of eqs.~\ref{eqn:wp-coefficient:neutral} and \ref{eqn:wp-coefficient:cation}
are replaced by:
\begin{align}
 C_{a\bfn a'\bfn'}^{\rm frozen}\left(t,t_0\right) & = \E^{\I E_\rmN \left(t-t_0\right) / \hbar}, & \nonumber \\
 D_{b'\bfm'b\bfm}^{\rm frozen}\left(t,t_1\right) & = \E^{\I E_\rmI \left(t-t_1\right) / \hbar}. & \nonumber
\end{align}
All other matrix elements and the initial nuclear wave packet, represented by
the $C_{a'\bfn'}\left(t_0\right)$ coefficients in Eq.~\ref{eqn:wp:neutral},
remain unchanged. As the result, this simulation can still exhibit isotope
effects due to the differences in the initial ground-state wave packet widths
combined with coordinate dependence of the matrix elements.
\item[``AC'':]
In this ``autocorrelation'' simulation, the electronic part of the dipole is
calculated exactly as in the frozen-nuclei simulation ``F''.  For each
trajectory, the dipole is multiplied by the magnitude of the nuclear
autocorrelation function (section~\ref{sec:results:auto}). Thus, electronic
continuum and vibronic dynamics are assumed to be factorized and mutually
independent. This is the the approach typically taken for the analysis of PACER
spectra.
\item[``ND'':]
In this ``no-dipole'' simulation, we use the full Eq.~\ref{eqn:dipole:final}.
However, the effects of the laser field are neglected when evaluating vibronic
matrix elements of Eqs.~\ref{eqn:wp-coefficient:neutral} and
\ref{eqn:wp-coefficient:cation}. Thus, vibronic dynamics in the neutral species
and the cation are taken to be field-free, and depend only on the initial
composition of the wave packet and the elapsed time. The laser field interacts
with the continuum part of the wavefunction at all times, but with the rest of
the molecule only at the moment of ionization, through the $\Upsilon_{b\bfm
a'''\bfn'''}$ matrix element of Eq.~\ref{eqn:ionization:final}.  Electronic and
nuclear dynamics are otherwise fully correlated.
\item[``D'':]
Finally, the ``dipole'' simulation uses Eq.~\ref{eqn:dipole:final}, with no
further simplifications to any of the matrix elements. The laser field can
cause (subcycle) vibrational excitation in the parent neutral species, modify
the nuclear wave packet through the ionization matrix elements $\Upsilon_{b\bfm
a'''\bfn'''}$, and cause vibronic transitions on the cationic surface. ``D'' is
our best-effort simulation.
\end{itemize}

\begin{figure}[thbp]
\includegraphics[width=8.26cm]{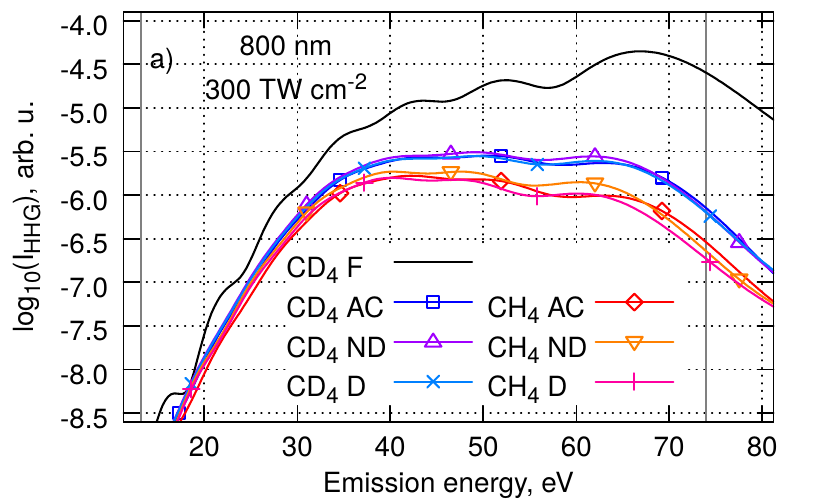} \\
\includegraphics[width=8.26cm]{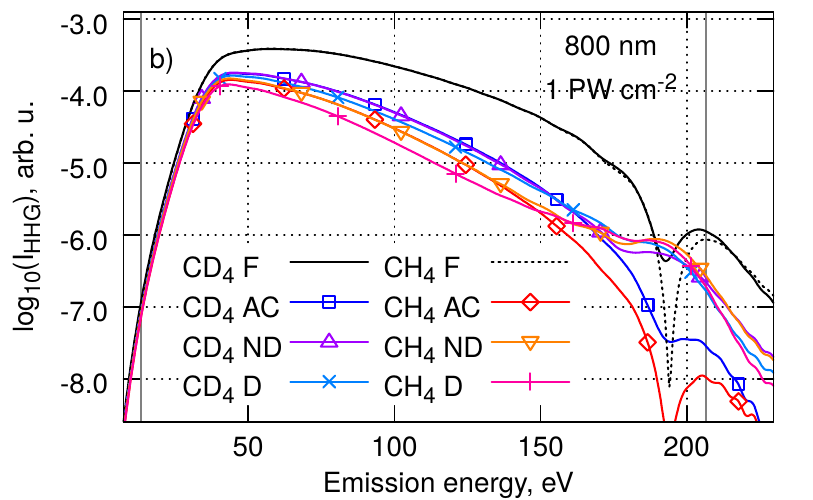}
\caption{\label{fig:harm-800} (Color online) MC-SFA-GWP high-harmonics spectrum of methane for
a single-sided recollision in $800$-nm driving field. Only short trajectories
are included. The ionization potential ($\ip$) and the harmonic cut-off
($3.17\up+1.3\ip$) are indicated by vertical grey lines. The labels are as
follows (see text for details): Nuclear geometry frozen, no nuclear factors
included: ``F'' ($\rm CD_4$: solid black line; $\rm CH_4$: dotted black line).
Nuclear geometry frozen, autocorrelation factor included: ``AC'' ($\rm CD_4$:
blue; $\rm CH_4$: red).  MC-SFA-GWP, vibronic terms due to laser coupling
omitted: ``ND'' ($\rm CD_4$: purple; $\rm CH_4$: orange).  MC-SFA-GWP, all
terms included: ``D'' ($\rm CD_4$: teal; $\rm CH_4$: magenta).  Panel a) 300
TW~cm$^{-2}$. The ``$\rm CD_4$~F'' and ``$\rm CH_4$~F'' (not shown) curves are
visually indistinguishable.  Panel b) 1 PW~cm$^{-2}$.
}
\end{figure}

Calculated normal and deuterated methane HHG spectra for the 800-nm driving
field are shown in Fig.~\ref{fig:harm-800}.  At the moderate,
$300$~TW~cm$^{-2}$ intensity of the driving field (panel a), harmonic cut-off
is found at $\approx 74$~eV photon energy (harmonic $47$). For these field
parameters, the frozen-nuclei spectra of the two isotopic species are
indistinguishable. Close to the cut-off, nuclear motion leads to HHG
suppression by a factor of $\approx 75\times$ ($\rm CD_4$) or $100\times$ ($\rm
CH_4$). All three approaches we consider for treating the nuclear motion yield
essentially the same results, indicating that the vibronic and recollision
dynamics are largely uncorrelated, and the sub-cycle vibronic dynamics is not
affected by the laser field.

The situation changes when the intensity is increased to $1$~PW~cm$^{-2}$
(Fig.~\ref{fig:harm-800}b). Now, the HHG cut-off extends to $\approx 206$~eV
(H$133$), past the minimum in the recombination cross-section (See
Fig.~\ref{fig:cross}). In the vicinity of the minimum, the finite width of the
initial nuclear wave packet now leads to isotope dependence even in the absence
of sub-cycle nuclear motion: HHG emission from $\rm CD_4$ (``$\rm CD_4$~F'',
solid black line) in the vicinity of the minimum is $\approx 55\times$ stronger
than for $\rm CH_4$ (``$\rm CH_4$~F'', dotted black line). At a first glance,
this result appears counter-intuitive: the broader ground-state vibrational
wave packet in $\rm CH_4$ is expected to sample a wider range of nuclear
configuration, and thus better ``fill in'' the $196$~eV structural minimum in
the recombination matrix element. However, in the vicinity of the $\rm CH_4$
harmonic minimum ($\approx 194$~eV), the phase of the recombination matrix
element varies substantially over the characteristic extent of the ground-state
nuclear wavefunction. The destructive interference then leads to a much
stronger suppression in ``$\rm CH_4$ F'' harmonic emission than might have been
expected from the matrix elements at the equilibrium geometry alone. The
narrower distribution of the nuclear positions in the heavier $\rm CD_4$
reduces the extent of the destructive interference, leading to increased HHG
intensity near the minimum.

The qualitative features of the frozen-nuclei high-harmonics spectra remain
essentially unchanged at longer wavelengths: the two isotopic species are
indistinguishable away from the structural minimum in the recombination matrix
elements. In the vicinity of the structural minimum, $\rm CD_4$ shows much
smaller suppression compared to the lighter isotopologue. As the result, we
will neither show nor discuss the frozen-nuclei HHG spectra for longer
wavelengths.

Returning to the $800$-nm, $1$~PW~cm$^{-2}$ case (Fig.~\ref{fig:harm-800}b),
the results obtained using different approaches for the treatment of sub-cycle
nuclear motion remain similar within the harmonic plateau (below $\approx
150$~eV), but start to differ closer to the structural minimum.  As expected,
the factorized ``AC'' approach faithfully reproduces the shape of the
frozen-nuclei spectra ($\rm CH_4$: solid red line; $\rm CD_4$: solid blue line,
Fig.~\ref{fig:harm-800}b), and yields a pronounced minimum around $194$~eV.
However, the minimum is completely filled-in when using a more elaborate
treatment (``ND'' or ``D''), reflecting the spreading of the nuclear wave packet
between ionization and recombination. The magnitude of the isotope effects is
also much reduced, and is possibly inverted (see
Section~\ref{sec:results:isotope} below). Furthermore, laser-field modification
of vibronic dynamics may begin to play a role.

\begin{figure}[thbp]
\includegraphics[width=8.26cm]{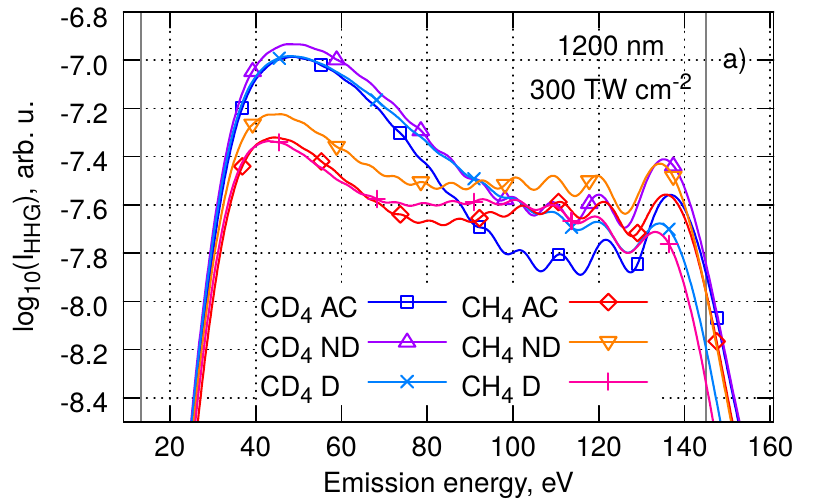} \\
\includegraphics[width=8.26cm]{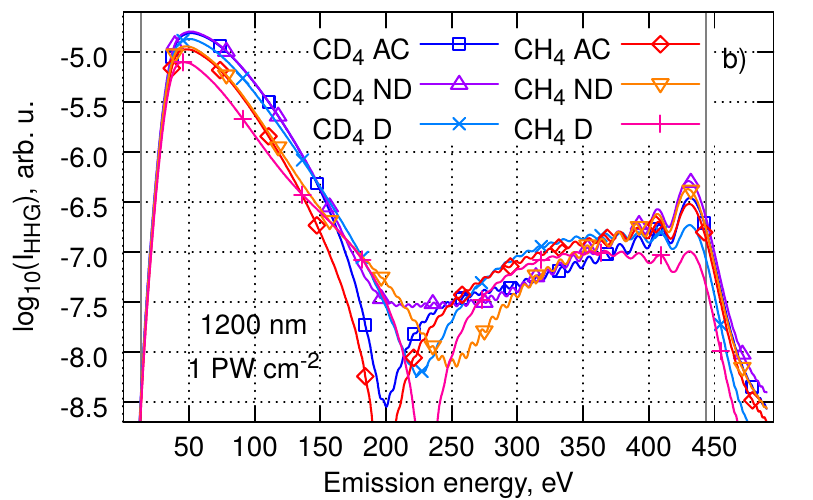}
\caption{\label{fig:harm-1200} (Color online) MC-SFA-GWP high-harmonics spectrum of methane in
$1.2$-$\mu$m driving field.  Panel a) 300 TW~cm$^{-2}$. Panel b) 1
PW~cm$^{-2}$. Also see Fig.~\ref{fig:harm-800} caption.  The HHG spectra for
the frozen nuclear configuration are off the scale, and are not shown.
}
\end{figure}

Calculated HHG spectra for single-sided recollision at $1.2$~$\mu$m are
collected in Fig.~\ref{fig:harm-1200}.  At $300$~TW~cm$^{-2}$, the cut-off is
found at $\approx145$~eV (H140), well below the structural minimum. On the
low-energy side of the plateau (below $\approx 80$~eV), the three approaches to
the treatment of nuclear motion (``AC'', ``ND'', and ``D'') yield very similar
results. This is not unexpected: this low-energy part of the spectrum
corresponds to recollision time delays below $1.7$~fs, which were explored by
the $800$-nm results above. At longer time delays, correlated vibronic
calculations (both ``ND'' and ``D'') now predict significantly higher HHG
intensity than the direct-product ``AC'' approximation (but still two orders of
magnitude lower than the frozen-nuclei results). The enhancement is due to the
coordinate dependence of the strong-field ionization
(Eq.~\ref{eqn:ionization:final}). Close to the peak of the laser electric
field, it leads to a substantial population of the the nuclear basis functions
excited along the $\nu_3$ (asymmetric stretch) and $\nu_4$ (asymmetric bend)
normal modes.  For $\rm CH_4$ and linear polarization along $X$
(Fig.~\ref{fig:dyson}), the initial amplitudes of the singly-excited $\nu_4$
($\nu_3$) on the dominant $D_3$ electronic surface reach $14\%$ ($33$\%) of the
vertical-ionization amplitude at $300$~TW~cm$^{-2}$, increasing to $14\%$
($36$\%) at 1~PW~cm$^{-2}$.  Ionization-induced nuclear wave packet reshaping is
smaller, but still substantial for $\rm CD_4$: $\nu_4$ ($\nu_3$) relative
amplitudes of $9\%$ ($19$\%) at $300$~TW~cm$^{-2}$, increasing to $10\%$
($21$\%) at 1~PW~cm$^{-2}$. The initially-excited component of the vibronic
wave packet reaches the half-revival faster (after $\approx 2$~fs delay), while
the higher magnitude of the revival (Fig.~\ref{fig:auto-vib}) compensates for
the reduced population relative to the vertical ionization. In the heavier $\rm
CD_4$, the $\nu_4$/$\nu_3$ half-revivals occur later (after $\approx 2.2$~fs),
so that the ``ND'' and ``AC'' spectra remain similar until much closer to the
cut-off. Finally, the ``ND'' (no laser coupling in the vibronic dynamics) and
``D'' (full laser coupling) spectra begin to diverge close to the cut-off,
indicating that field-induced bound-state dynamics becomes important beyond
$2$~fs delays.

At the higher 1~PW~cm$^{-2}$ intensity (Fig.~\ref{fig:harm-1200}b), the
harmonics cut-off now extends to $443$~eV (H429). The structural minimum in the
recombination matrix elements is now well within the plateau region, inducing a
false cut-off near $200$~eV. The direct product form (``AC'') remains a
reasonable approximation early within the plateau (up to $\approx1.5$~fs;
$160$~eV emission energy).  At higher emission energies, both the correlations
between the vibronic and continuum dynamics and the field-induced vibronic
transitions become important. The position of the structural minimum is shifted
to higher photon energies ($225$--$250$~eV, depending on the species and the
details of the treatment). A similar shift of an HHG feature was experimentally
observed\cite{Baker08a} in PACER experiments on $\rm H_2$, where however coordinate
dependence of the recombination matrix elements moves the minimum to
\textit{lower} energies.

Vibronic dynamics ``fills in'' the structural minimum (``$\rm CD_4$ ND'':
purple line with triangles; ``$\rm CH_4$ ND'': orange line with inverted
triangles).  At the same time, field-induced vibronic transitions sharpen the
minimum again (``$\rm CD_4$ ND'': teal line with crosses; ``$\rm CH_4$ ND'':
pink line with plus signs).  The isotope effects in the high-energy part of the
plateau thus reflect a number of dynamics (see \ref{sec:results:isotope}
below).

\begin{figure}[thbp]
\includegraphics[width=8.26cm]{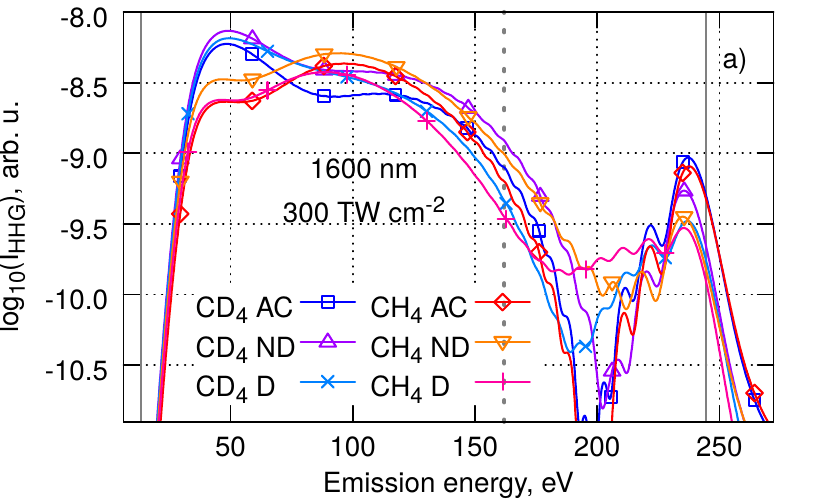} \\
\includegraphics[width=8.26cm]{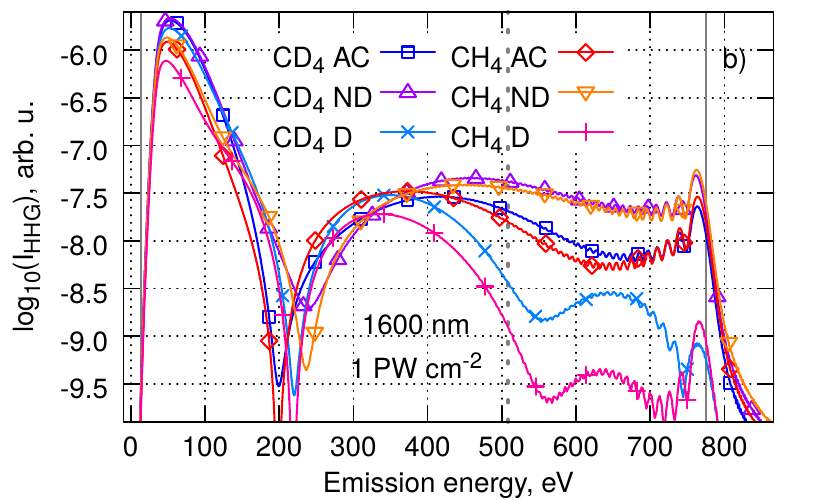}
\caption{\label{fig:harm-1600} (Color online) MC-SFA-GWP high-harmonics spectrum of methane in
$1.6$-$\mu$m driving field.  Panel a) 300 TW~cm$^{-2}$. Panel b) 1
PW~cm$^{-2}$. Also see Fig.~\ref{fig:harm-800} caption.  The HHG spectra for
the frozen nuclear configuration are off the scale, and are not shown.  Parts
of the ``ND'' and ''D'' spectra between the dotted vertical grey line and the
cut-off may show increased computational errors, see
Section~\ref{sec:details:ac}.
}
\end{figure}

Our final example uses a $1.6$-$\mu$m driver (Fig.~\ref{fig:harm-1600}). Already
at $300$~TW~cm$^{-2}$, (panel a) the cut-off extends beyond the structural
minimum, to $244$~eV (H315). The qualitative features of the calculated HHG
spectra are similar to the $1.2$-$\mu$m, 1~PW~cm$^{-2}$ case above: the
simplified product representation (``AC'') is adequate at low energies (short times);
correlated vibronic dynamics and laser-induced vibronic coupling are
increasingly important at higher energies, and especially in the vicinity of
the structural minimum. At the $1$~PW~cm$^{-2}$ intensity, the still higher harmonic
cut-off ($775$~eV, H1000), leads to a particularly clean demonstration of a transition
from largely pure, field-free vibronic dynamics to fully-correlated dynamics
under strong-field control (Fig.~\ref{fig:harm-1600}b). On the left side of the
plateau (before $170$~eV/$1.6$~fs), the spectra are grouped by the isotopic
species, with all three treatments (``AC'', ``ND'', and ``D'') yielding very
similar results for either species.

On the right-hand side of the plateau (beyond $450$~eV/$2.3$~fs), the HHG
spectra are instead grouped by the treatment of bound--continuum correlations
and laser field effects in vibronic dynamics. For example, at $600$~eV, the
product form (``$\rm CD_4$~AC'', blue line with squares) underestimates the HHG
intensity by a factor of $\approx 3\times$ compared the field-free correlated
treatment of the vibronic dynamics (``$\rm CD_4$~ND'', purple line with
triangles). At the same time, the full treatment including field-induced
vibronic transitions (``$\rm CD_4$~D'', teal line with crosses) predicts HHG
yield at this energy $\approx 2.3\times$ \textit{below} the ``AC'' treatment,
and $\approx 8\times$ below the field-free dynamics prediction.  For nuclear
dynamics simulations neglecting field-induced vibronic transitions, isotopic
effects in this region do not exceed $1.5\times$. However, the full treatment
(``$\rm CH_4$~D'', purple line with plus signs) predicts a very strong
suppression of harmonic emission beyond $500$~eV, which is likely to appear as
a false cut-off in experiment. Clearly, neglecting field-induced bound-state
transitions in the cation is not a viable treatment in the high-intensity
and multi-cycle limit\cite{Chirila08a,Castiglia11a}.

Finally, in the transition region around the structural minimum, both the
position and the depth of the structural minimum are highly sensitive to the
isotopic species, treatment of non-adiabatic dynamics, and laser-induced
bound-state dynamics. Again similar to the $1.2$-$\mu$m, 1~PW~cm$^{-2}$ case,
correlated vibronic dynamics shifts the apparent position of the structural
minimum to higher energies and reduces its contrast, while inclusion of the
laser-induced vibronic dynamics partially reverses the shift. In $\rm CH_4$,
the minimum is found at $198$~eV (``AC''), $236$~eV (``ND''), and $219$~eV
(``D'').  In $\rm CD_4$, the minima are shifted slightly to higher photon
energies: $200$~eV (``AC''), $239$~eV (``ND''), and $220$~eV (``D'').

\subsection{Isotope effects\label{sec:results:isotope}}

Because PACER experiments are routinely interpreted in terms of
ionization-recollision time delays, it is instructive to examine the isotopic
ratios of the high-harmonics spectra in
Figs.~\ref{fig:harm-800}--\ref{fig:harm-1600} in a similar way. In the absence
of continuum resonances, there exists a one-to-one mapping between the
short-trajectory HHG spectrum and the ionization-recollision time
delay\cite{Corkum93a}. This relationship breaks down close to sharp features in
photorecombination matrix elements (See Section~\ref{sec:results:hhg} and
Refs.~\cite{Smirnova09b,Patchkovskii12a,Smirnova13a,Smirnova13b}).  Although the emission
time in these cases can still be recovered through the time-frequency
analysis\cite{Chirila10a}, there appears to be no unambiguous way of
reconstructing the ionization-recollision time delay close to resonances. We
therefore choose the classical ``simple man's'' mapping\cite{Corkum93a} (SMM) for
the ionization--recombination time delay, with the twin caveats: a) this
mapping is known to be inaccurate for short trajectories, especially for the
$800$-nm driver\cite{Chirila10a,Haessler11a}; and b) the mapping should be
treated as undefined close to structural minima in the harmonic spectrum. The
advantage of the SMM is that the range of possible time delays depends only on
the wavelength of the driving laser, so that different intensities can be
compared directly. The underlying assumption of the PACER method is that the
isotope effects represented in this form are intensity- and
wavelength-independent.

\begin{figure}[thbp]
\includegraphics[width=8.26cm]{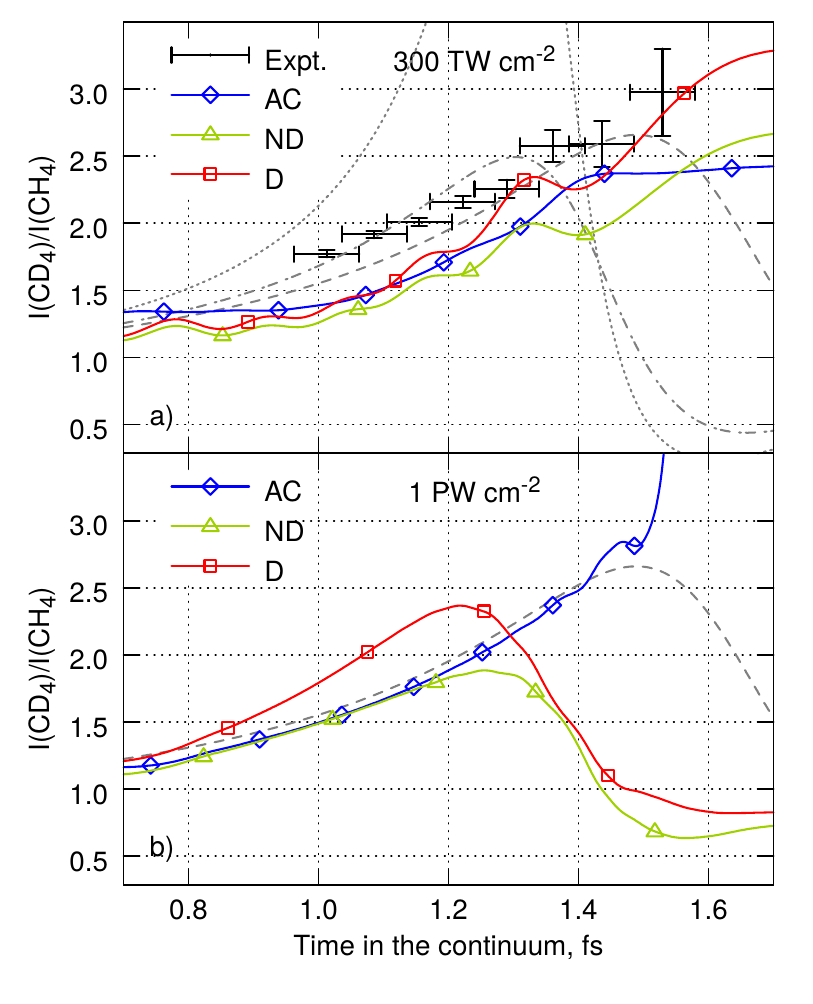}
\caption{\label{fig:iso-800} (Color online) Isotope effects in methane for the $800$-nm driving
field.  Vertical axis: The ratio of the calculated HHG yields for $\rm CD_4$ and
$\rm CH_4$ (Fig.~\ref{fig:harm-800}).  Horizontal axis: Time delay between
ionization and recollision events, calculated from the harmonic energy using
classical simple man's model (see text). ``AC'' (blue line with diamonds):
direct product of the autocorrelation function and frozen-nuclei electron
dynamics; ``ND'' (green line with triangles): correlated treatment,
field-induced bound-state dynamics neglected; ``D'' (red line with squares):
the full treatment, including field-induced vibronic transitions. Dashed grey
line gives the ratio of ground-state autocorrelation factors
(Fig.~\ref{fig:auto}). Dash-dotted and dotted grey lines in panel a) represent
the ratio of autocorrelation factors for $\nu_3$ and $\nu_4$
vibrationally-excited initial wave packets (Fig.~\ref{fig:auto-vib}),
respectively.  Panels a) and b) correspond to laser intensity of $300$ and
$1000$~TW~cm$^{-2}$, respectively. Experimental points in panel a) are from
Ref.~\cite{Baker06a}, measured at the estimated intensity of $200$~TW~cm$^{-2}$.
}
\end{figure}

Calculated isotope effects for the $800$-nm driving field are shown in
Fig.~\ref{fig:iso-800}.  The available experimental data\cite{Baker06a} (black
error bars) was obtained at the estimated intensity of $200$~TW~cm$^{-2}$, and
can be most directly compared to the numerical results of panel (a), calculated
at a somewhat higher intensity of $300$~TW~cm$^{-2}$. As expected from the
similarity of the calculated spectra for the three approximations we consider
here (Fig.~\ref{fig:harm-800}a), the calculated PACER ratios are nearly
identical. The simple ratio of the autocorrelation functions
(Fig.~\ref{fig:auto}) yields nearly identical results, except very close to the
harmonic cut-off ($t>1.5$~fs). The agreement with the experiment is
satisfactory, although the experimental isotope ratios are consistently
slightly higher than the calculated values. A possible reason for the
discrepancy is the vibrational excitation of the neutral molecule by the
raising edge of the $8$~fs pulse used in experiment\cite{Baker06a}, which is
not included in the present single-cycle simulation.  As can be seen from
Fig.~\ref{fig:auto-vib}, population of the IR-active $\nu_3$ and $\nu_4$
vibrational modes in the initial wave packet is expected to increase isotope
effects for the delays below $1.3$~fs (dash-dotted and dotted lines in Fig.~\ref{fig:iso-800}a).

Both the intrinsic correlations and laser-induced vibronic dynamics become
important for the isotope effects at the higher, $1$~PW~cm$^{-2}$ intensity
(Fig.~\ref{fig:iso-800}b). Compared to the uncoupled approximation (``AC'',
blue line with diamonds), correlations between the continuum and vibronic
dynamics in the cation (``ND'', green line with triangles) substantially reduce
the calculated isotope effects.  Laser-induced vibronic dynamics (``D'', red
line with squares) partially counteracts this effect. Our best-effort
calculation (``D'') suggests that the observed isotope effects at this
intensity should start decreasing beyond $1.2$~fs delays, with inverse isotope
effects predicted beyond $1.45$~fs. We emphasize that the inverse
isotope-effect in this case is due to the structural minimum at $196$~eV in our
photoionization matrix elements (Fig.~\ref{fig:cross}) and the associated
breakdown of the time-frequency mapping. An inverse isotope effect of a similar
origin was previously predicted for the $\rm D_2$/$\rm H_2$
pair\cite{Chirila08a,Chirila09b}.  A change in the position of the photoionization resonance
will also change the apparent time delay where the inverse isotope effect is
predicted. 

\begin{figure}[thbp]
\includegraphics[width=8.26cm]{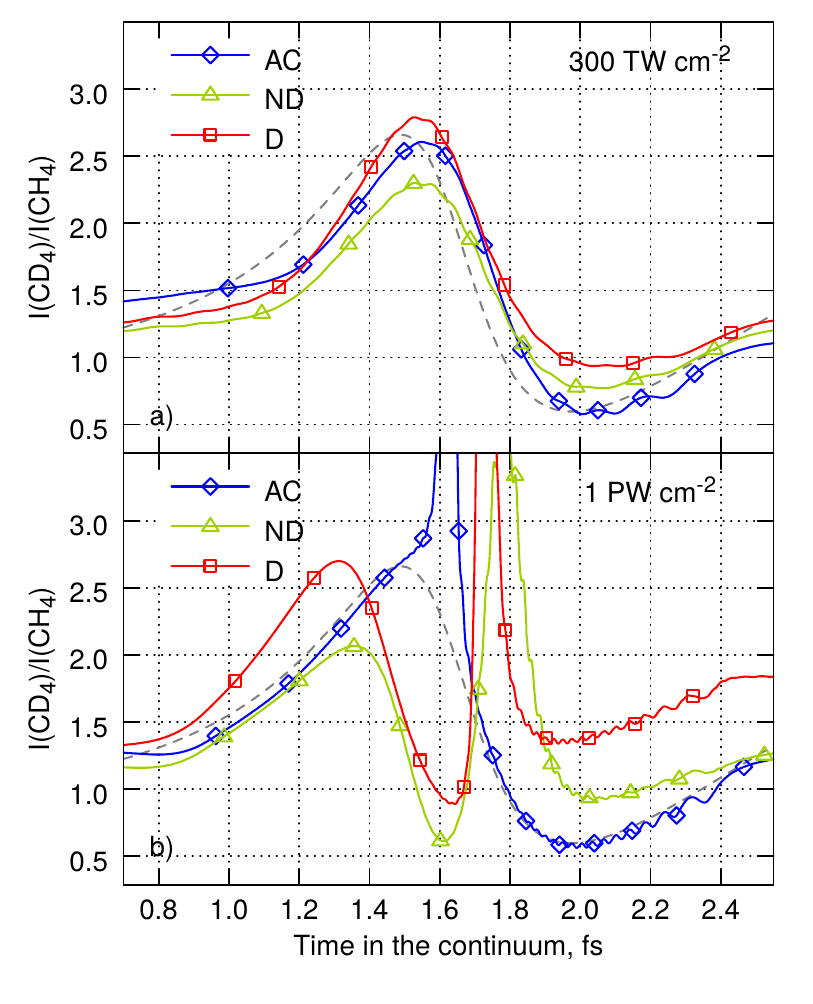}
\caption{\label{fig:iso-1200} (Color online) Isotope effects in methane for $1.2$-$\mu$m
driving field, from HHG data in Fig.~\ref{fig:harm-1200}. Also see
Fig.~\ref{fig:iso-800} caption.
}
\end{figure}

The simulated PACER results for the longer-wavelength, $1.2\mu$m driver and
moderate $300$~TW~cm$^{-2}$ intensity are shown in Fig.~\ref{fig:iso-1200}a.
Superficially, this PACER trace is remarkably similar to the ``D'' (correlated
and laser-coupled) result at $800$~nm and $1$~PW~cm$^{-2}$
(Fig.~\ref{fig:iso-800}b): the isotope effect first increases, then changes
sense at longer delay times. However, the physics behind the trace is entirely
different. At $1.2\mu$m, the signal reflects the field-free vibronic dynamics
in the transient cation, with neither vibronic-continuum correlations nor
laser-driven vibronic dynamics qualitatively affecting the result. Thus,
neglecting the laser coupling in the cation (``ND'') and neglecting both the
laser coupling and vibronic-continuum correlations (``AC'') yield results very
similar to the full simulation. All three simulated PACER traces are nearly
on-top of the simple ratio of the autocorrelation functions (dashed line). At
the $1.2\mu$m wavelength, the laser cycle is long enough to allow vibronic
wave packet in the cation to reach the half-revival at the
conical-intersection point (See Fig.~\ref{fig:auto} and Section~\ref{sec:results:auto}).  
The PACER trace
containing a reversal of the isotope effect thus represents a true signature of
the CI dynamics.

The situation becomes more complex at the higher $1$~PW~cm$^{-2}$ intensity
(Fig.~\ref{fig:iso-1200}b).  Now, the harmonic spectrum extends far enough to
access the structural minimum in the recombination matrix elements. As the
result, the apparent isotope effects at time delays corresponding to the
structural minimum (``AC'': $\approx 1.63$~fs, ``ND'': $\approx 1.78$~fs,
``D'': $\approx 1.73$~fs) become very large, and sensitive to the details of
the treatment (``AC'': $14\times$; ``ND'': $3.9\times$; ``D'': $175\times$). At
time delays unaffected by the structural minimum ($t<1.55$~fs and $t>1.85$~fs),
the calculated PACER signal is qualitatively similar to the $300$~TW~cm$^{-2}$
results. As was already seen above for the $800$~nm case, the influence of
continuum-vibronic correlations and laser-driven vibronic dynamics increases at
the higher intensity. 

\begin{figure}[thbp]
\includegraphics[width=8.26cm]{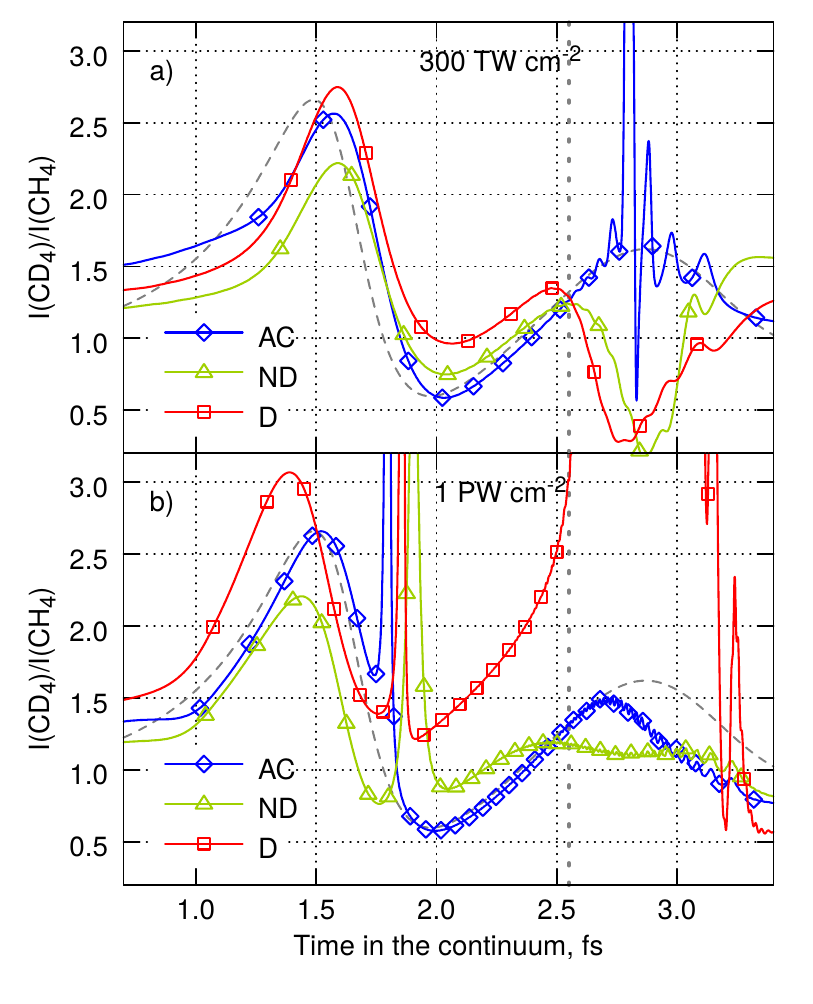}
\caption{\label{fig:iso-1600} (Color online) Isotope effects in methane for $1.6$-$\mu$m
driving field, from HHG data in Fig.~\ref{fig:harm-1600}. Time delays beyond
the grey vertical dotted line at $2.6$~fs may suffer from increased numerical 
errors, see Section~\ref{sec:details:ac} of the text.  Also see 
Fig.~\ref{fig:iso-800} caption.
}
\end{figure}

Finally, simulated PACER traces at $1.6\mu$m wavelength are shown in
Fig.~\ref{fig:iso-1600}.  Already at the moderate $300$~TW~cm$^{-2}$ intensity,
harmonic cutoff reaches beyond the structural minimum. The corresponding PACER
trace (panel a) is representative of the intrinsic vibronic dynamics for the delays below
$\approx2.5$~fs. At longer times, the structural-minimum feature
dominates. Further increasing the intensity (panel b) leads to a very complex
spectrum, which loses nearly all information on the intrinsic vibronic dynamics
beyond $\approx 1.7$~fs.  If the intrinsic, field-free dynamics of the cation
is of interest, extending the wavelength to $1.6\mu$m therefore appears to
offer little advantage.

\section{Conclusions and outlook\label{sec:conclusions}}

In this work, we introduced MC-SFA-GWP -- a version of the molecular strong-field
approximation which treats all electronic and nuclear degrees of freedom,
including their correlations, quantum-mechanically.  The new technique allows,
for the first time, realistic simulations of the nuclear motion effects on
high-harmonic emission in polyatomic molecules without invoking
reduced-dimensionality models for the nuclear motion or the electronic
structure.

We use the new technique to model isotope effects in methane.  The intermediate
$^2F_2$ electronic state of the $\rm CH_4^+$ cation transiently accessed by the
HHG process possesses a symmetry-required triple-state conical intersection at
the Franck-Condon point.  Simulations of the field-free vibronic dynamics of
the $^2F_2$ state in $\rm CH_4^+$ indicates that a fraction of the
initially-prepared wave packet undergoes a half-revival, accompanied by a sign
change of the vibronic wavefunction, within $2.2$~fs. This time scale is well
within the laser-cycle duration of near-IR light, and is accessible with HHG
spectroscopy.  The revival is associated with population of an intermediate
vibronic wave packet, composed of the singly-excited $\nu_4$ ($t_2$) and $\nu_2$
($e$) vibrational modes coupled to the degenerate components of the $^2F_2$
electronic surface. In the space of nuclear coordinates, this wave packet forms
a prolate spheroidal shell around the conical intersection. The revival times
are determined predominantly by the strength of the vibronic coupling near the CI, thus
permitting its direct experimental measurement.

A well-understood difficulty in PACER and other HHG spectroscopies of molecules
is the strong (Gaussian in time) suppression of the HHG emission due to the
loss of nuclear wavefunction overlap. We demonstrate that beyond $1.5$~fs,
nuclear motion in methane no longer causes a Gaussian suppression of the HHG
signal, with the expected nuclear factor persisting at $\approx 1$\% level up
to $3.5$~fs -- high enough to allow experimental detection. This unexpected
persistence of the fraction of the HHG signal appears to be universal, and
has been predicted for several other cations\cite{Forster13a,Arnold17a}.

We analyze and identify a number of physical mechanisms which contribute to the
isotopic PACER signal in methane. At intensities below $300$~TW~cm$^{-2}$ and
wavelengths below $1.2\mu$m, the autocorrelation contribution is dominant. This
contribution derives from the intrinsic vibronic dynamics around the conical
intersection. It manifests as an inverse isotope effect at $\approx2.2$~fs
ionization-recollision delay, and is wavelength- and intensity-independent.
This effect has been predicted previously\cite{Mondal14a,Mondal15a}. An
experimental observation was claimed very recently\cite{Lan17a}.

The autocorrelation term is modified by a number of wavelength- and/or
intensity-dependent contributions, including: coordinate dependence of
strong-field ionization amplitudes\cite{Chelkowski96a}; laser-driven vibronic
dynamics in the neural and the transient cation\cite{Castiglia11a}; the
coordinate dependence of the photorecombination
amplitudes\cite{Baker08a,Chirila08a,Chirila09b}. For wavelengths beyond
$\approx1.6\mu$m or intensities above $300$~TW~cm$^{-2}$, the latter
contributions may dominate the PACER spectrum (see also
Ref.~\cite{Chirila08a,Chirila09b}), even for a single-cycle driving pulse.

Furthermore, we demonstrate that the PACER concept breaks down for harmonic
emission close to resonances (constructive or destructive) in the recombination
matrix elements.  At these energies, no simple relationship exists between the
harmonic photon energy and the emission time. The isotopic ratio is then no
longer representative of the intrinsic vibronic dynamics in the cation. For any
given combination of the wavelength and intensity of the driving field, the
PACER spectrogram resulting from a resonance may be indistinguishable from the
signal due to the intrinsic vibronic dynamics. It is therefore essential to
perform PACER experiments at a number of wavelengths and/or intensities.
Features which appear at a fixed, or nearly fixed emission energy are
likely to originate from a resonance in the recombination matrix elements. 

Although the present analysis focuses on one of the simplest polyatomic
molecules, the methane, both conical intersections and resonant features in
photorecombination matrix elements are ubiquitous in polyatomic molecules.
High-harmonics spectroscopy thus provides a tool for exploring both, in the
regime not easily accessible with other techniques. 

The effects discussed presently are of the \textit{sub-cycle} nature, and are
already present for a single-cycle driving pulse. For longer pulses, additional
physical mechanisms will come into
play\cite{Chelkowski96a,Daniele09a,Castiglia11a,Morales14a,Lara-Astiaso16a}.  A careful examination of
these multi-cycle effects and their interaction with the intrinsic sub-cycle
dynamics is one of the possible future applications of the MC-SFA-GWP approach.


\appendix

\section{Evaluation of the $d t_1$ integral in Eq.~\ref{eqn:dipole:int-k}\label{sec:appendix-A}}

We need to evaluate an integral in the form:
\begin{align}
  I & = \int_{t_0}^{t} d t_1 f\left(\veck_\rms\left(t_1\right)\right) 
        \E^{-\I \phi_d\left(\veck_\rms\left(t_1\right),t,t_1\right)}, \label{eqn:appa:integral}
\end{align}
where $\veck_\rms$ is in turn a function of $t_1$ (Eq.~\ref{eqn:stationaryk})
and $\phi_d$ is defined by Eq.~\ref{eqn:dipole:phase}.  We assume that
$f\left(\veck_\rms\right)$ depends at most linearly on $\veck_\rms$. Expanding
$\phi_d$ through the \textit{third}\cite{Ivanov96a} order in $t_1$ around $t_\rms$,
changing integration variable to $\tau=t_1-t_\rms$, and extending the
integration limits to infinity, we obtain:
\begin{widetext}
\begin{align}
  I & \approx \int_{-\infty}^{\infty} d \tau \left[ f 
               + \tau \frac{\partial f}{\partial \veck_\rms} \cdot \frac{\partial\veck_\rms}{\partial t_\rms} 
               + \tau \frac{\partial f}{\partial t_\rms} \right]
        \E^{-\I \phi_d - \I \frac{\partial \phi_d}{\partial t_\rms} \tau 
                       - \frac{\I}{2} \frac{\partial^2 \phi_d}{\partial t_\rms^2} \tau^2 
                       - \frac{\I}{6} \frac{\partial^3 \phi_d}{\partial t_\rms^3} \tau^3}, \label{eqn:appa:integral-expanded} \\
  \frac{\partial^3 \phi_d}{\partial t_\rms^3} &= - \frac{3\hbar}{m} \frac{\veck_\rms^2}{\left(t-t_\rms\right)^2} 
                                                 + \frac{3 e}{m} \frac{1}{t-t_\rms} \veck_\rms \cdot \vecF
                                                 + \frac{e}{m} \veck_\rms \cdot \frac{\partial}{\partial t_\rms} \vecF
                                                 - \frac{e^2}{m\hbar} \vecF^2, \label{eqn:appa:d3phidts3}
\end{align}
\end{widetext}
\begin{align}
  \frac{\partial\veck_\rms}{\partial t_\rms} & = \frac{\veck_\rms}{t-t_\rms} - \frac{e}{\hbar} \vecF, \label{eqn:appa:dksdts} \\
  \frac{\partial \phi_d}{\partial t_\rms} &= - \frac{\hbar}{2 m} \veck_\rms^2 - \frac{1}{\hbar} \ip, \label{eqn:appa:dphidts} \\
  \frac{\partial^2 \phi_d}{\partial t_\rms^2} &= - \frac{\hbar}{m} \frac{\veck_\rms^2}{t-t_\rms} 
                                                 + e \veck_\rms \cdot \vecF. \label{eqn:appa:d2phidts2}
\end{align}
In Eq.~\ref{eqn:appa:integral-expanded}, $f$, $\phi_d$, their derivatives, and
$\vecF$ are evaluated at $\veck_\rms$ and $t_\rms$ pairs solving
Eq.~\ref{eqn:stationaryts}.  For linearly-polarized driving field,
$\frac{\partial^2 \phi_d}{\partial t_\rms^2}$ (Eq.~\ref{eqn:appa:d2phidts2})
and all but the last term in $\frac{\partial^3 \phi_d}{\partial t_\rms^3}$
(Eq.~\ref{eqn:appa:d3phidts3}) vanish. For low-frequency fields, these terms
remain negligible for moderate non-zero ellipticities as well (high harmonic
signal vanishes for large ellipticities\cite{Ivanov96a}), so that:
\begin{align}
  \frac{\partial^2 \phi_d}{\partial t_\rms^2} &\approx 0, & \tag{\ref{eqn:appa:d2phidts2}a} \\
  \frac{\partial^3 \phi_d}{\partial t_\rms^3} &\approx - \frac{e^2}{m\hbar} \vecF^2. & \tag{\ref{eqn:appa:d3phidts3}a}
\end{align}

We now note that (Eq.~10.4.32 of Ref.~\cite{Abramowitz72a}):
\begin{align}
  \int_{-\infty}^{\infty}   \E^{\I a t^3 + \I x t} d t &= 
            \frac{2\pi}{\left(3 a\right)^{1/3}} \Ai\left(\frac{x}{\left(3 a\right)^{1/3}}\right), \label{eqn:appa:airy}
\end{align}
\begin{align}
  \int_{-\infty}^{\infty} t \E^{\I a t^3 + \I x t} d t &= 
           -\I\frac{2\pi}{\left(3 a\right)^{2/3}} \Ai'\left(\frac{x}{\left(3 a\right)^{1/3}}\right), \label{eqn:appa:airyprime}
\end{align}
where $\Ai$ is an Airy function.  (Eq.~\ref{eqn:appa:airyprime} is obtained by
differentiating Eq.~\ref{eqn:appa:airy} with respect to $x$.) Therefore,
Eq.~\ref{eqn:appa:integral-expanded} becomes:
\begin{widetext}
\begin{align}
  I &\approx \E^{-\I \phi_d} 2 \pi \Bigg[
              f \left(\frac{2 m\hbar}{e^2 \vecF^2}\right)^{1/3} \Ai\left(\zeta\right) 
           - \I \left(\frac{\partial f}{\partial t_\rms} 
           + \frac{\partial f}{\partial \veck_\rms} \cdot \frac{\partial\veck_\rms}{\partial t_\rms} \right)
              \left(\frac{2 m\hbar}{e^2 \vecF^2}\right)^{2/3} \Ai'\left(\zeta\right),
    \Bigg] \label{eqn:appa:final}
\end{align}
\end{widetext}
\begin{align}
  \zeta & = \left(\frac{2m}{e^2\hbar^2\vecF^2}\right)^{1/3}\left(\ip + \frac{\hbar^2\veck_\rms^2}{2m}\right), \label{eqn:appa:zeta}
\end{align}
where again all quantities are evaluated at pairs $\veck_\rms$, $t_\rms$
satisfying Eq.~\ref{eqn:stationaryts}. If multiple roots are present, summation
over all roots is implied.

\section{Derivation of Eq.~\ref{eqn:ionization:final}\label{sec:appendix-B}}

The presence of the vibronic state energy expectations $E_{b\bfm}$ and
$E_{a'''\bfn''}$, rather than the manifold-average $\ip$, in
Eq.~\ref{eqn:ionization:argument} is intuitively appealing, but requires some
additional justification.  Application of Eq.~\ref{eqn:appa:final} to the
integral of Eq.~\ref{eqn:dipole:int-k} leads to:
\begin{widetext}
\begin{align}
 \Upsilon_{b\bfm a'''\bfn'''} & = 
     \vecF\!\left(t_\rms\right) \cdot \vecR_{b\bfm a'''\bfn'''}\left(\veck_\rms\right) 
     2 \pi  \left(\frac{2 m}{e^2 \hbar^2 \vecF^2\!\left(t_\rms\right)}\right)^{1/3} \Ai\left(\zeta\right) \nonumber \\
    & - 2\I\pi \vecF\!\left(t_\rms\right) \cdot \vecR_{b\bfm a'''\bfn'''}\left(\veck_\rms\right) 
        \frac{\partial}{\partial t_s} \log C_{a'''\bfn'''}\left(t_\rms\right)
        \left(\frac{2 m}{e^2 \hbar^2\vecF^2\!\left(t_\rms\right)}\right)^{2/3} \Ai'\left(\zeta\right) \nonumber \\
    & - 2\I\pi \vecF\!\left(t_\rms\right) \cdot \vecR_{b\bfm a'''\bfn'''}\left(\veck_\rms\right) 
        \frac{\partial}{\partial t_s} \log D_{b'\bfm'b\bfm}\left(t,t_\rms\right)
        \left(\frac{2 m}{e^2 \hbar^2\vecF^2\!\left(t_\rms\right)}\right)^{2/3} \Ai'\left(\zeta\right) \nonumber \\
    & - 2\I\pi  \frac{\partial}{\partial \veck_\rms} 
             \left[ \vecF\!\left(t_\rms\right) \cdot \vecR_{b\bfm a'''\bfn'''}\left(\veck_\rms\right) \right]
        \cdot \left( \frac{\hbar\veck_\rms}{t-t_\rms} - e \vecF\!\left(t_\rms\right) \right)
              \left(\frac{2 m}{e^2 \hbar^2\vecF^2\!\left(t_\rms\right)}\right)^{2/3} \Ai'\left(\zeta\right),
    \label{eqn:ion:intermediate}
\end{align}
\end{widetext}
where we have neglected the $t_\rms$ and $\veck_\rms$ dependence of the
recombination dipole and the overall prefactor and $\zeta$ is given by Eq.~\ref{eqn:appa:zeta}.
Formal differentiation of Eqs.~\ref{eqn:wp-coefficient:neutral}--\ref{eqn:wp-coefficient:cation} gives:
\begin{widetext}
\begin{align}
  \frac{\partial}{\partial t_1} D_{b'\bfm'b\bfm}\left(t,t_1\right) & = 
      \frac{\I}{\hbar} \E^{\I E_\rmI \left(t-t_1\right) / \hbar} 
      \bra{\bfm'} \bra{X_{b'}} 
         \hat{U}_\rmI\left(t,t_1\right) \left[\hat{H}_\rmI\left(t_1\right) - E_\rmI\right]
      \ket{X_{b}}\ket{\bfm},
      \label{eqn:d-derivative} \\
  \frac{\partial}{\partial t_1} C_{a\bfn a'\bfn'}\left(t_1,t_0\right) & = 
      - \frac{\I}{\hbar} \E^{\I E_\rmN \left(t_1-t_0\right) / \hbar} 
      \bra{\bfn} \bra{\Phi_{a}} 
         \left[\hat{H}_0\left(t_1\right) - E_\rmN\right]
         \hat{U}_0\left(t_1,t_0\right) 
      \ket{\Phi_{a'}}\ket{\bfn'}.
      \label{eqn:c-derivative}
\end{align}
\end{widetext}
If the Hamiltonians $\hat{H}_\rmI$ and $\hat{H}_0$ are diagonally-dominant in
the basis of corresponding vibronic product states,
Eqs.~\ref{eqn:d-derivative}--\ref{eqn:c-derivative} reduce to:
\begin{align}
  \frac{\partial}{\partial t_1} D_{b'\bfm'b\bfm}\left(t,t_1\right) &\approx  
      \frac{\I}{\hbar} \left( E_{b\bfm} - E_\rmI\right) 
      D_{b'\bfm'b\bfm}\left(t,t_1\right), \label{eqn:d-derivative:simple} \\
  \frac{\partial}{\partial t_1} C_{a\bfn a'\bfn'}\left(t_1,t_0\right) &\approx
     -\frac{\I}{\hbar} \left( E_{a\bfn} - E_\rmN\right)
      C_{a\bfn a'\bfn'}\left(t_1,t_0\right), \label{eqn:c-derivative:simple}
\end{align}
where $E_{b\bfm}$ and $E_{a\bfn}$ are given by 
Eqs.~\ref{eqn:ion-state-energy}--\ref{eqn:neutral-state-energy}.

Substituting Eqs.~\ref{eqn:d-derivative:simple}--\ref{eqn:c-derivative:simple}
into Eq.~\ref{eqn:ion:intermediate}, we note that the first three terms are in
fact the lowest-order contributions to the Taylor expansion of the first term
in Eq.~\ref{eqn:ionization:final} for $\delta E = \left(E_{b\bfm} -
E_{a'''\bfn'''}\right)-\ip$.  Similar expansion for the second term corresponds
to a term quadratic in $\tau$, neglected in deriving Eq.~\ref{eqn:appa:final}.
To within the accuracy expected from Eq.~\ref{eqn:ionization:final}, we can
therefore replace $\ip$ by $\left(E_{b\bfm} - E_{a'''\bfn'''}\right)$ in the
last term of Eq.~\ref{eqn:ion:intermediate} as well, giving
Eq.~\ref{eqn:ionization:argument}. The exponential part of
Eq.~\ref{eqn:ionization:argument} coincides with the result of the weak-field
asymptotic tunneling theory (WFAT)\cite{Tolstikhin17a}.

\section{Evaluation of the Fourier-transform integrals in eq.~\ref{eqn:ionization:dipole}\label{sec:appendix-C}}

Evaluation of matrix elements appearing in Eq.~\ref{eqn:ionization:dipole}
requires calculation of Fourier transforms of the ``cradle'' orbitals
$\vec{\phi}^\rmC_{b a}$ and products of the Dyson orbitals $\phi^\rmD_{b a}$
and a dipole operator $\vecr$.  Here, both Dyson and ``cradle'' orbitals are
given by an expansion over atom-centered Cartesian Gaussian-type orbitals.  The
desired integrals are readily obtained by a simple modification of standard
1-electron integral packages. 

Indeed, closely following the approach of Ahlrichs\cite{Ahlrichs06a}, the
primitive integral $I_{\bf0,\bf0}$ is given by (cf. Eq.~7 of
\cite{Ahlrichs06a}):
\begin{align}
  I_{\bfzero,\bfzero}   
        & = \braopket{\bfzero}{\E^{\I \veck\cdot\vecr}}{\bfzero} \nonumber \\
        & = \int d\vecr \E^{-\alpha\left|\vecr-\bfA\right|^2} \E^{\I \veck\cdot\vecr} 
                \E^{-\beta\left|\vecr-\bfB\right|^2} \nonumber \\
        & = \E^{\xi\left|\bfB-\bfA\right|^2} \left(\frac{\pi}{\zeta}\right)^{3/2} 
                \E^{\I \veck\cdot\bfP} \E^{-\frac{\veck^2}{4\zeta}}, \label{eqn:appb:i0}
\end{align}
\begin{align}
  \ket{\bfa} &= \left(x-A_x\right)^{a_x} \left(y-A_y\right)^{a_y} \left(z-A_z\right)^{a_z} 
                \E^{-\alpha\left|\vecr-\bfA\right|^2}, \label{eqn:appb:basis} \\
  \bfP  & = \frac{\alpha\bfA+\beta\bfB}{\alpha+\beta}, \label{eqn:appb:P} \\
  \zeta & = \alpha + \beta, \label{eqn:appb:zeta} \\
  \xi   & = \frac{\alpha\beta}{\alpha+\beta}. \label{eqn:appb:xi}
\end{align}
Functions $\ket{\bfa}$ are unnormalized Cartesian Gaussians with quantum
numbers $\bfa$ and exponent $\alpha$, centered at $\bfA$ (Eq.~1 of
\cite{Ahlrichs06a}), and analogously for $\ket{\bfb}$.  Applying the usual
generating operators ($\hat{M}$ of Eqs.~17 and 29 of \cite{Ahlrichs06a}), we
immediately obtain the recursion relation:
\begin{align}
 I_{\bfa+\bfone_p,\bfb}    &=
       \left(\bfP_p - \bfA_p + \frac{\I k_p}{2\zeta}\right) I_{\bfa,\bfb} \nonumber \\
   & + \frac{a_p}{2\zeta} I_{\bfa-\bfone_p,\bfb} 
     + \frac{b_p}{2\zeta} I_{\bfa,\bfb-\bfone_p}, \label{eqn:appb:s-recursion}
\end{align}
where $p$ is a Cartesian direction ($p=x,y,z$), $\bfa=(a_x,a_y,a_z)$, and
$\bfone_p$ is a unit 3-vector containing $1$ in position $p$.  Noting that:
\begin{align}
  r_p \ket{\bfa} &= \ket{\bfa+\bfone_p} + A_p \ket{\bfa}, \label{eqn:appb:dipole}
\end{align}
we then obtain:
\begin{align}
    \braopket{\bfa}{r_p\E^{\I \veck\cdot\vecr}}{\bfb} &=
           I_{\bfa+\bfone_p,\bfb} + A_p I_{\bfa,\bfb} \nonumber \\
   &=  \left(\bfP_p + \frac{\I k_p}{2\zeta}\right) I_{\bfa,\bfb} \nonumber \\
   & + \frac{a_p}{2\zeta} I_{\bfa-\bfone_p,\bfb} 
     + \frac{b_p}{2\zeta} I_{\bfa,\bfb-\bfone_p}. \label{eqn:appb:d-recursion}
\end{align}

We note that Eqs.~\ref{eqn:appb:s-recursion} and \ref{eqn:appb:d-recursion} are
nearly identical to the standard recursion relations for the overlap and dipole
integrals \cite{Obara86a,Obara88a}.  Finally, by choosing: $\bfb = \bfzero$, $\beta =
0$, and $\bfB = \bfzero$ in Eqs.~\ref{eqn:appb:s-recursion} and
\ref{eqn:appb:d-recursion}, we obtain the desired expressions for the Fourier
transforms of the primitive Gaussian $\ket{\bfa}$ and its first Cartesian
moments.


\end{document}